\documentclass[conference]{IEEEtran}
\usepackage{cite}
\usepackage{graphicx}
\usepackage{subfig}
\usepackage{multirow}
\usepackage{booktabs}
\usepackage{algorithm}
\usepackage{algpseudocode}
\usepackage{amsfonts}
\usepackage{url}
\usepackage{flushend}
\usepackage[acronym,nomain,nonumberlist]{glossaries}

\makeglossaries

\newacronym{rpi}{RPI}{Raspberry Pi 3B+}
\newacronym{tcp}{TCP}{Transmission Control Protocol}
\newacronym{udp}{UDP}{User Datagram Protocol}
\newacronym{tls}{TLS}{Transport Layer Security}
\newacronym{ws}{WS}{WebSocket}
\newacronym{quic}{QUIC}{Quic}
\newacronym{wss}{WSS}{WebSocket Secure}
\newacronym{mqtt}{MQTT}{Message Queuing Telemetry Transport}
\newacronym{gso}{GSO}{Generic Segmentation Offload}
\newacronym{aoi}{AoI}{Age of Information}
\newacronym{iot}{IoT}{Internet of Things}
\newacronym{tc}{TC}{Traffic Control}
\newacronym{nsa}{NSA}{Non Standalone}
\newacronym{rssi}{RSSI}{Received Signal Strength Indicator}
\newacronym{rsrp}{RSRP}{Reference Signals Received Power}
\newacronym{lte}{LTE 4G}{Long Term Evolution}
\newacronym{ng}{NG 5G}{New Generation}
\newacronym{gps}{GPS}{Global Positioning System}
\newacronym{nat}{NAT}{Network Address Translation}
\newacronym{ntp}{NTP}{Network Time Protocol}
\newacronym{mno}{MNO}{Mobile Network Operator}
\newacronym{lan}{LAN}{Local Area Network}
\newacronym{usb}{USB}{Universal Serial Bus}
\newacronym{uart}{UART}{Asynchronous Receiver-Transmitter}
\newacronym{rtt}{RTT}{Round Trip Time}
\newacronym{fifo}{FIFO}{First In First Out}
\newacronym{lifo}{LIFO}{Last In First Out}
\newacronym{http}{HTTP}{Hypertext Transfer Protocol}
\newacronym{rsrq}{RSRQ}{Reference Signals Received Quality}
\newacronym{ttl}{TTL}{Time to Live}
\newacronym{ip}{IP}{Internet Protocol}
\newacronym{rdt}{RDT}{Reliable Data Transfer}
\newacronym{arq}{ARQ}{Automatic Repeat reQuest}
\newacronym{crc}{CRC}{Cyclic Redundancy Check}
\newacronym{qos}{QOS}{Quality of Service}
\newacronym{mac}{MAC}{Media Access Control}
\newacronym{csma}{CSMA}{Carrier Sense Multiple Access}
\newacronym{api}{API}{Application programming interface}
\newacronym{icmp}{ICMP}{Internet Control Message Protocol}
\newacronym{ber}{BER}{Bit Error rate}
\newacronym{toml}{TOML}{Tom's Obvious, Minimal Language}
\newacronym{mtu}{MTU}{Maximum Transmission Unit}
\newacronym{mss}{MSS}{Maximum Segment Size}
\glsaddall

\begin{document}
%
\title{Age of Information and Energy Consumption in IoT: an Experimental Evaluation}
%
%
%

\author{\IEEEauthorblockN{Federico Cristofani}
\IEEEauthorblockA{University of Pisa, Italy \\
f.cristofani@studenti.unipi.it}
\and
\IEEEauthorblockN{Valerio Luconi}
\IEEEauthorblockA{IIT-CNR, Italy \\
valerio.luconi@iit.cnr.it}
\and
\IEEEauthorblockN{Alessio Vecchio}
\IEEEauthorblockA{University of Pisa, Italy \\
alessio.vecchio@unipi.it}
}

\maketitle

\begin{abstract}
The Age of Information (AoI) is an end-to-end metric frequently used to understand how ``fresh'' the information about a remote system is. In this paper, we present an experimental study of the relationship between AoI and the energy spent by the device that produces information, e.g. an IoT device or a monitoring sensor. Such a relationship has been almost neglected so far, but it is particularly important whenever the sensing side is battery-operated. The study is carried out in a scenario where access is achieved via the cellular network and information is transferred using MQTT, a popular messaging protocol in the IoT domain. Numerous parameters of operation are considered, and the most efficient solutions in all configurations are provided.
\end{abstract}


\IEEEpeerreviewmaketitle

\section{Introduction}
Age of Information (AoI) has emerged as a critical metric reflecting the timeliness and relevance of data delivery. In a monitoring system, AoI is defined as the time elapsed since the generation of the latest received packet at the monitor, and it can capture the freshness of the received data better than classic metrics such as delay~\cite{Abbas2023:comprehensive}. In today's interconnected world, where real-time data processing and transmission are paramount, understanding and optimizing the AoI is pivotal for a spectrum of applications, including multimedia streaming, Internet of Things (IoT) deployments, and financial transactions. Since its introduction~\cite{Kaul2012:real}, AoI has been used as a metric to analyze and characterize a wide variety of systems, especially IoT-related such as intelligent transportation systems, vehicular networks, smart agriculture, and augmented reality~\cite{patra2016minimizing, suma2017iot, Chaccour2020:ruin}.

Alongside the quest for timely information delivery, the energy footprint of network protocols and infrastructure has gained significant attention. With the exponential growth of digital traffic, data centers, and communication networks, the environmental impact of energy-intensive operations has become a focal point for researchers, policymakers, and industry stakeholders alike. Especially in the IoT context, energy consumption considerations gain paramount importance, since most of the devices are battery-operated~\cite{Caiazza2022:saving, Caiazza2022:edge, Caiazza2023:measuring, Caiazza2024:energy}.

Recently, various studies have investigated the trade-off between maintaining a desired freshness level of information while minimizing energy consumption in IoT systems, mainly from a theoretical perspective~\cite{Tripathi2017:age, Valehi2017:maximizing}. Conversely, recent experimental works still have not focused on the implication of AoI on energy but were mainly aimed at studying AoI-aware systems in some operating scenarios~\cite{Beytur2020:towards, Kadota2021:wifresh}.

In this work, we aim to evaluate the relationship between AoI and energy consumption in an experimental publish-subscribe scenario. We built a testbed made of an IoT device that publishes updates at a given rate, and we measure AoI at a subscriber. Simultaneously, we measure energy consumption via a hardware power monitor on the IoT device. The devices are connected via a cellular network and communicate through the MQTT protocol on top of QUIC and TLS. We run an extensive set of experiments to explore how a wide range of parameters of operations can influence the two considered metrics. In detail, we considered the rate of generation of messages on the IoT device, the latency between the IoT device and the MQTT broker, the messages' payload size, the computational capacity of the IoT device, and the impact of the underlying transport protocol. Thanks to the visualization instrument of Pareto fronts, we point out the most efficient solutions for every configuration of parameters. We show that there is no clear indication of a universally optimal solution, but the context and the system's requirements should indicate which of the most efficient solutions to adopt. However, we believe that our experimental results are extremely precious in helping researchers and practitioners in future protocol designs, IoT system architectures, and environmental sustainability initiatives.

The rest of the paper is structured as follows. Section~\ref{sec:relwork} provides an overview of the existing literature on the subject. Section~\ref{sec:setup} describes our experimental setup. In Section~\ref{sec:metrics}, we describe the metrics and the aggregation techniques that we use to analyze experimental results. Section~\ref{sec:results} shows our findings. Finally, Section~\ref{sec:conclusion} concludes the paper.
\section{Related Work}
\label{sec:relwork}

The performance of transport and application protocols in an \acrshort{iot} context has been extensively studied from various points of view. Several works have evaluated the adoption of the recently introduced \acrshort{quic} transport protocol in \acrshort{iot}. In~\cite{Eggert2020TowardsST} the authors demonstrated the feasibility of a \acrshort{quic} standard implementation designed for tiny devices, by quantifying storage, computing, memory, and energy requirements, however, without involving any application protocol. Other studies have used \acrshort{http} as an application protocol over \acrshort{quic} in an \acrshort{iot} environment. In~\cite{http-quic-iot}, the authors studied the latency and scalability of HTTP/3 over \acrshort{quic} in an \acrshort{iot} system in the cloud/edge continuum, showing results in the order of hundreds or thousands of milliseconds depending on the load of the system. The energetic performance of \acrshort{http} protocols have been studied for both \acrshort{iot} devices and smartphones in~\cite{Caiazza2023:measuring, Caiazza2024:energy}, where the authors compared different versions of the \acrshort{http}, showing that the ones using \acrshort{quic} as an underlying protocol consumed more energy. The authors also showed that their results could be due to the current lack of maturity and standardized adoption of \acrshort{quic} implementations, among other factors. Fatima et al. compared \acrshort{quic} and \acrshort{tcp}/\acrshort{tls} as underlying protocols of \acrshort{mqtt}, for what concerns latency, in a simulated environment. They showed that, for short-lived connections, the performance of the two protocols is comparable, while for long-lived connections, \acrshort{quic} obtains shorter completion times. The same comparison is conducted in~\cite{quic-iot-gateway} for a ]\acrshort{iot}-cloud environment based on \acrshort{mqtt}. Again, \acrshort{quic} obtained a much better performance. In~\cite{energy-quic-mqtt}, the authors analyzed \acrshort{mqtt} over \acrshort{quic} and \acrshort{tcp}/\acrshort{tls} from the energetic point of view, showing that for small-medium delays \acrshort{quic} was able to consume less energy, while for large ones \acrshort{tcp}/\acrshort{tls} obtained a slightly better performance.

Work on AoI focused on theoretically analyzing the performance of different systems characterized by different queuing models and scheduling algorithms in different application scenarios~\cite{Kaul2011:minimizing, Yates2012:real, Inoue2018:analysis, Chiariotti2021:quic}.

The relationship between energy and AoI has been investigated mainly from a theoretical point of view. In~\cite{Tripathi2017:age}, the authors devise two scheduling algorithms for multiple sensors that minimize the AoI at the monitoring station. Subsequently, energy consumption is brought into play and the authors show variants to their algorithms that can achieve close to optimal performance with significant energy consumption reduction. A framing and scheduling policy for minimizing energy consumption in cognitive wireless sensor networks has been proposed in~\cite{Valehi2017:maximizing}. The optimization is conducted under strict constraints of AoI. Via numerical and simulation analysis, the authors provide the optimal number of samples per packet under given operational conditions, such as the number of nodes in the network and channel quality indicators. In~\cite{Gu2019:timely}, the authors derive closed forms for AoI and energy consumption in an IoT monitoring system based on low power wide area (LPWA) wireless communication technologies and adopting a truncated automatic repeat request (TARQ) scheme. The authors show that under transmit power constraint, the TARQ scheme obtains lower AoI than a classic ARQ scheme. The problem of minimizing energy consumption under AoI constraints has been studied also in~\cite{Saurav2023:online}, where the authors derived a lower bound on the competitive ratio of any given causal policy for choosing which packets to transmit on a given node. In addition, they propose a greedy policy that achieves that lower bound. Other works have tackled the problem of AoI and energy from different perspectives, for example for an energy harvesting source~\cite{Bacinoglu2018:achieving}, or computation offloading in industrial IoT~\cite{Huang2023:aoi}.

Few works focused on evaluating AoI in experimental environments. In~\cite{Sonmez2018:age}, the AoI of a TCP/IP connection over WiFi, Ethernet, and cellular networks has been investigated via emulation and in a physical network. Observations have been conducted with different sampling rates and degree of network load. The AoI of TCP and UDP connections over the Internet and of lightweight IoT connections in a local WiFi network has been evaluated in~\cite{Beytur2020:towards}. In this work, the authors provide insight into how energy can be a factor impacting AoI in constrained environments such as IoT. In~\cite{Kadota2021:wifresh}, the authors proposed WiFresh: an AoI-aware wireless architecture to achieve optimal AoI even in the case of an overloaded network. They show that compared to classic WiFi, WiFresh can obtain lower values of AoI.

The analysis of the previous works reveals considerable attention to topics related to the \acrshort{iot}, highlighting extensive coverage of these issues. In particular, the emerging protocol as \acrshort{quic} is gaining popularity as a modern alternative to the de-facto standard \acrshort{tcp} that showed limitations for lossy networks and constrained devices. The focus is also placed on the \acrshort{aoi} metric, able to represent the freshness of data, that in modern systems assumes a pivotal role. However, significant gaps emerged. Among all, there is a lack of experimental works, especially regarding combined studies on \acrshort{aoi} and energy. This work is placed at the intersection of studies exploring energy consumption and \acrshort{aoi} in \acrshort{iot}, adopting an experimental methodology. To the best of our knowledge, ours is the first experimental work that analyzes \acrshort{aoi} and energy consumption in a real-world \acrshort{iot} scenario.
\section{Experimental Setup}
\label{sec:setup}

\begin{figure}[t!]
    \centering
    \includegraphics[width=0.85\columnwidth]{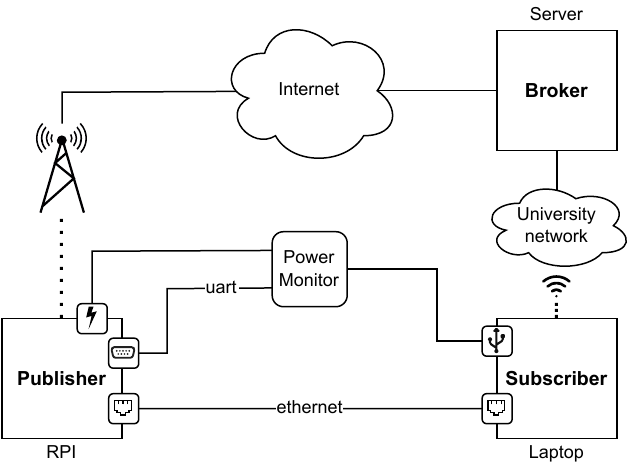}
    \caption{Experimental setup hardware and software architecture.}
    \label{fig:setup_overview}
\end{figure}

\begin{table*}[!t]
    \centering
    \caption{Experimental setup hardware specs.}
    \begin{tabular}{llll}
        \toprule\toprule
        \multicolumn{1}{c}{\textbf{Device}} & \multicolumn{1}{c}{\textbf{CPU}} & \multicolumn{1}{c}{\textbf{RAM}} & \multicolumn{1}{c}{\textbf{NIC}} \\
        \midrule
        \acrshort{rpi} & BCM2837B0 ARMv8 64bit\footnote{The processor has four cores, but only one is enabled to reproduce a resource-constrained device faithfully} & 1GB LPDDR2 & SIM8200EA-M2 5G HAT \\
        \midrule[0pt]
        Laptop & Intel Core i7-8565U & 16GB DDR4 & Intel(R) Wireless-AC 9260 \\
        \midrule[0pt]
        Server & Intel Xeon Gold 5120\footnote{The server is a virtual machine that can benefit from only 2 cores out of a total of 14 available} & 4GB & Ethernet adapter \\
        \bottomrule
        \multicolumn{4}{c}{\small {\textit {Ethernet interfaces for \acrshort{rpi} and the laptop are not reported in the table}}}
    \end{tabular}
    \label{tab:hardware}
\end{table*}

\begin{table}
    \centering
    \caption{Network connections in the experimental setup.}
    \begin{tabular}{lrrrr}
    \toprule\toprule
        \multicolumn{1}{c}{\textbf{Connection}} & 
        \multicolumn{1}{c}{\textbf{\acrshort{rtt}}} & 
        \multicolumn{1}{c}{\textbf{Download}} & 
        \multicolumn{1}{c}{\textbf{Upload}} &
        \multicolumn{1}{c}{\textbf{Hops}} \\ 
    \midrule
    \acrshort{rpi} - Server & 80 ms & 28 Mbit/s & 15 Mbit/s & 12 \\
    \midrule[0pt]
    Laptop - Server & 5 ms & 200 Mbit/s & 160 Mbit/s & 3 \\
    \bottomrule
    \end{tabular}
    \label{tab:net_stats}
\end{table}

This study aims to evaluate \acrfull{aoi} and the energy consumption in a \acrshort{iot} scenario, using the \acrshort{mqtt} protocol running on top of multiple transport protocols: \acrshort{quic} and \acrshort{tcp}/\acrshort{tls}. We built the experimental setup depicted in Figure~\ref{fig:setup_overview}, composed of multiple hardware and software components. The \acrshort{rpi} acts as an \acrshort{iot} device, hosting the client side, i.e. the publisher in the \acrshort{mqtt} terminology. The \acrshort{rpi} is equipped with a cellular network card~\cite{cellular-hat} able to support both \acrfull{lte} and \acrfull{ng} communication technologies\footnote{Experiments have been carried out only using 4G access.}. The \acrshort{rpi} represents a battery-powered device. To measure its power consumption, it is powered by the Otii Arc Pro from Qoitech power monitor~\cite{otii}. The Otii power monitor is featured with a \acrfull{uart} interface that is used to send commands to the \acrshort{rpi} device and to annotate the recorded trace with timestamps to precisely confine the portion of the total energy required during the execution of an experiment. The \acrshort{rpi} publisher is then connected via cellular connection to a \acrshort{mqtt} broker hosted on a server in the University of Pisa cloud network.

The final component of the experimental setup is a laptop, which has a twofold function. Firstly, it acts as a \acrshort{mqtt} subscriber client, connected to the broker via the University of Pisa network (with WiFi access). The connection between the broker and the subscriber is characterized by a relatively low latency, being the two components in the same network. Secondly, the laptop acts as the controller of the experiments. It is connected to both the \acrshort{rpi} and the Otii to provide the configuration parameters and the automation for each experiment.

The goal of the study is to evaluate the system from the point of view of the AoI and the energy needed on the publisher side. We are not interested in the energy consumption of the machine hosting the subscriber as we suppose that the alerting/controlling system mentioned before is not executed on a battery-operated device. Similarly, the energy needed to run the broker is not of interest, as it is generally executed on reasonably powerful machines without constraints in terms of energy. With this setup, we can measure both the energy consumption on the \acrshort{rpi} and the \acrshort{aoi}, as the difference between the instant when the message was generated and the instant when it was received. To obtain a precise measure of the \acrshort{aoi}, the two devices (the \acrlong{rpi} and the laptop) need to be adequately synchronized. To achieve the desired synchronization, we used an approach based on the \acrfull{ntp} protocol. The two devices are connected via a second ethernet connection, and the laptop acts as the time-reference server for the \acrshort{rpi}. The ethernet connection is characterized by an extremely low latency, being the two devices in the same lab room, and is used only for synchronization traffic. All the other traffic exchanged by the two devices flows via the cellular connection and the Internet.

The overall configuration corresponds to a rather common scenario in the \acrshort{iot} domain, in which the \acrshort{rpi} acts as a device installed somewhere that produces data, typically collecting information by monitoring the environment. The subscriber on the laptop in turn can be considered as a cloud application that deals with the collection of the data produced, applying logic that may be sensitive to the freshness of the information obtained from the data. The two parties exchange information through the broker, hosted in the cloud as the subscriber, which represents a centralized entity enabling indirect communication according to the \acrshort{mqtt} protocol. Table~\ref{tab:hardware} summarizes the hardware specs of the experimental setup, while Table~\ref{tab:net_stats} provides a characterization of the network connections between the hardware components.

\subsection{Implementation}
The client has been implemented relying on the Quinn library~\cite{quinn}, a rust-based implementation of the IETF \acrshort{quic} protocol. We did not use a standard \acrshort{mqtt} client as available \acrshort{quic}-based implementations are rather limited in number. Quinn provides an async API and it is based on the Tokio async runtime~\cite{tokio}. Quinn also uses the Rustls~\cite{rustls} library for the cryptographic functionality.  In particular, our client also uses the mqttbytes~\cite{mqttbytes} libraries for producing \acrshort{mqtt} messages according to the specification. The client just implements the subset of the \acrshort{mqtt} protocol needed for our experiments, like sending messages to the broker and receiving the corresponding acknowledgments. To achieve higher message rates, the client uses a task to send messages to the broker and another task to keep track of the corresponding incoming acks. The client also includes the buffering mechanisms previously mentioned: a task produces new information at a nominal rate and uses a buffer to communicate with the sender task. The latter is responsible for extracting information from the buffer and sending it through the network. The buffer is managed according to two strategies: (i) a FIFO queue with a capacity of 1, 16, and 1024 messages, and (ii) a drop-head-on-full queue of capacity 1. In the first case, the two tasks can be blocked when the queue is full or empty and no messages are lost. In the second case, a message can be lost if a new one is produced before the current one is transmitted. In this case, the producing task never blocks and always operates according to the nominal rate. The strategies will be hereafter indicated as FIFO 1, FIFO 16, FIFO 1024, and DROP.
The broker is an instance of the Rumqttd server. The subscriber is rather simple and receives messages when relayed by the broker. The subscriber computes the \acrshort{aoi}, by using a timestamp collected when the message is received and another timestamp included in the message by the publisher and indicating the time when the message was produced. The subscriber has been implemented using the Paho library~\cite{paho}.

\section{Collected Metrics and Aggregation}
\label{sec:metrics}
To make the paper self-contained, we here provide a description of the metrics and the aggregation tools that we consider in this study.

\subsection{Age of Information}
The recording of generation and reception instants of the message published during the time window gives all the information needed to compute the \acrshort{aoi} not only in correspondence with the message arrival but in any instant along the entire duration of the experiment, obtaining an exact form for the sawtooth shape function that characterizes the metric~\cite{yates2021age}. Dealing with an analytic expression of the function, rather than a discrete group of points, e.g. by sampling the time evolution of \acrshort{aoi}, allows for precise calculations of the statistical quantities, needed to aggregate the collected data for the analysis. The drawback is that those calculations might be not straightforward to implement. The statistical quantities taken into consideration are mean and median values.
The formula applied to compute the exact value of the \acrshort{aoi} in each instant is the following:
\begin{equation}
    \Delta(t)=t - U(t)
\end{equation}
where $\Delta(t)$ is the \acrshort{aoi} at instant $t$, and $U(t)$ the generation time of the newest data received. This formula represents the time evolution of the \acrshort{aoi} metric, which assumes the shape of an irregular sawtooth, rising linearly while waiting for new messages and suddenly falling to a lower value at the reception of a new message. The function is always positive, with a lower bound given by the one-way delay of the network. Each linear section has a $45^\circ$ inclination, i.e. slope 1, so the distance covered on the x-axis is the same as the one on the y-axis for each segment. The function is integrable.

\begin{figure}[!t]
    \centering
    \includegraphics[width=\columnwidth]{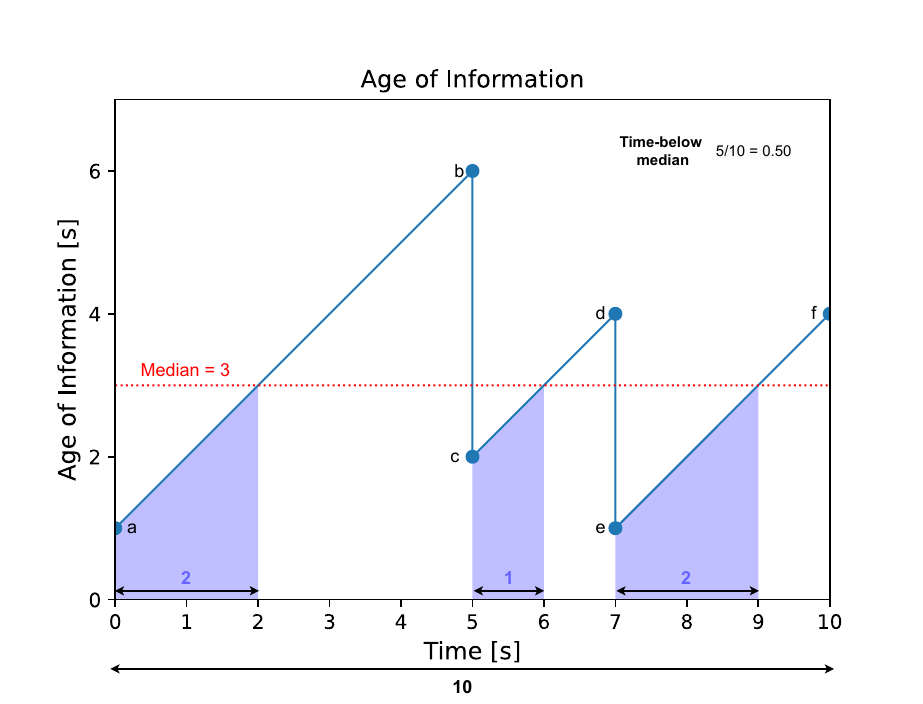}
    \caption{Median definition for \acrshort{aoi} function}
    \label{fig:aoi-median-definition}
\end{figure}

\noindent\textbf{Mean Value.} The mean value is the simplest form of data aggregation. However, dealing with a function requires a different approach than the simple arithmetic mean applicable for a discrete set of points. To compute the mean of the \acrshort{aoi} function we used the integral mean, defined as the ratio between the definite integral of the function over the given interval and the length of the interval:
\begin{equation}
    \begin{aligned}
        &f : [a, b] \rightarrow\mathbb{R}\textit{ bounded and integrable in } [a,b] \\
        &M(f,[a,b])=\frac{1}{b-a} \int_{a}^{b} f(x) \, dx
    \end{aligned}
\label{eq:integral-mean}
\end{equation}
In our case, the interval length is given by the time elapsed between the first and last message received by the subscriber.

\begin{figure}[!t]
    \centering
    \includegraphics[width=\columnwidth]{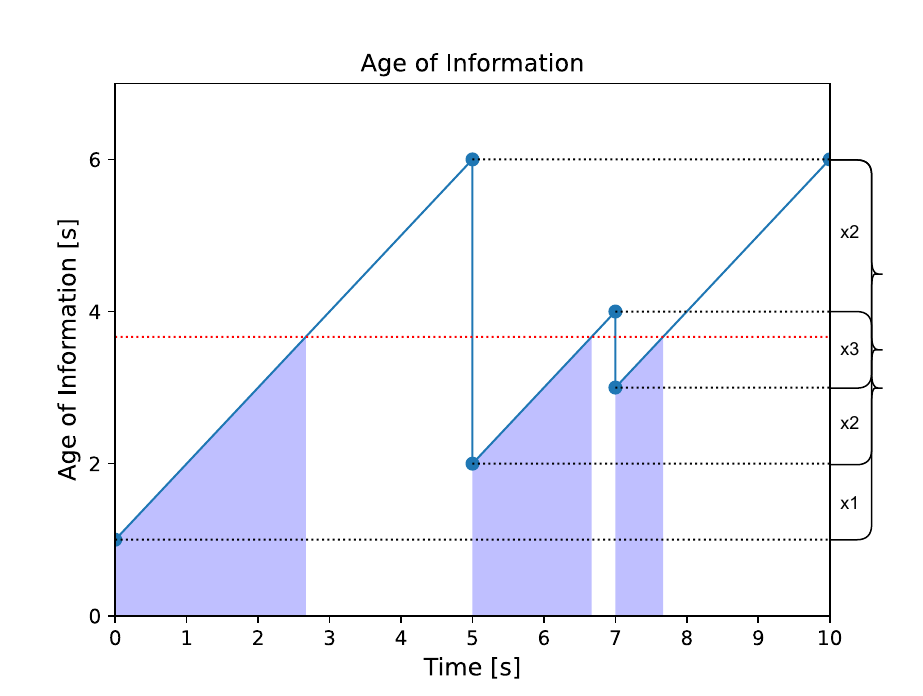}
    \caption{Projection algorithm.}
    \label{fig:projection-algorithm}
\end{figure}

\noindent\textbf{Median Value.} Besides the mean value, we compute the median value, which is known to be more robust to noise. The problem is to find an extension of the median applicable to a function and not to a discrete set of points, obtaining the same type of relationship that exists between the arithmetic mean and integral mean. The median is defined as the value separating the higher half from the lower half of the points in the set. The concept of half the points of the set can be extended to half the length of the interval over which the \acrshort{aoi} function is defined. Based on this new definition, the median value can be assumed as the level of \acrshort{aoi} for which the sum of the lengths of the intervals on the x-axis that have associated \acrshort{aoi} less than (greater than) the median is exactly half of the total length of the interval over which the function is defined. The visualization of this definition is reported in Figure~\ref{fig:aoi-median-definition}. To compute the median of a sawtooth function, we devised a method based on projecting the function segments on an axis. This method obtains the exact median value without involving complex computations. The exploited property is the unitary slope, which allows us to work on either the x-axis or the y-axis. The method is based on the projection of each segment on a vertical axis, referred to as the projection axis. The projections on the projection axis can generate overlapping regions. For each region, a weight is assigned, computed as the number of different projections that fall in it. Next, we compute a weighted sum as the length of each region times the associated weight. This sum will be referred to as the total extension. Finally, we iterate over the regions, bottom-up, computing their extension, i.e. their weighted length, and summing them up until half of the total extension is reached. The last sum will probably exceed half of the total extension, so only a fraction of the weighted length of the last region is to be considered. The point on which the iteration stops, i.e. the quote on the axis, corresponds to the median value. The construction of the projection axis is reported in Figure~\ref{fig:projection-algorithm} and a possible implementation is shown in Algorithm~\ref{alg:median}.

\begin{algorithm}
\caption{Projection median}
\label{alg:median}
\begin{algorithmic}
    \Function{compute\_projection\_median}{$\textit{timestamps}$} \Comment{Returns a float}
        \State $\textit{projections} \gets \text{empty list}$
        \State $\textit{extension} \gets 0$

        \For{$i \gets 1$ to $\text{len}(\textit{timestamps}) - 1$}
            \State $\textit{start\_aoi} \gets \textit{timestamps}[i - 1]["rx"] - \textit{timestamps}[i - 1]["gen"]$
            \State $\textit{end\_aoi} \gets \textit{timestamps}[i]["rx"] - \textit{timestamps}[i - 1]["gen"]$
            \State $\textit{projections.append}((\textit{start\_aoi}, 1))$ \Comment{+1 marks the begin of the projection}
            \State $\textit{projections.append}((\textit{end\_aoi}, -1))$ \Comment{-1 marks the end of the projection}
            \State $\textit{extension} \gets \textit{extension} + (\textit{end\_aoi} - \textit{start\_aoi})$
        \EndFor

        \State $\textit{projections.sort}(\text{key} \gets x: x[0])$ \Comment{Sort list of tuples using the second element as key}

        \State $\textit{weight} \gets 0$
        \State $\textit{curr\_ext} \gets 0$
        \State $\textit{start} \gets \text{-1}$

        \For{$\textit{projection} \text{ in } \textit{projections}$}
            \If{$\textit{start} \neq -1$}
                \State $\textit{region} \gets \textit{projection[0]} - \textit{start}$ \Comment{Region length}
                \If{$\textit{curr\_ext} + \textit{region} \times \textit{weight} \geq \textit{extension} / 2$}
                    \State \textbf{return} $\textit{start} + (\textit{extension} / 2 - \textit{curr\_ext})$
                \EndIf
                \State $\textit{curr\_ext} \gets \textit{curr\_ext} + \textit{region} \times \textit{weight}$
            \EndIf
            \State $\textit{weight} \gets \textit{weight} + \textit{projection[1]}$ \Comment{Weight is increased/decreased by 1}
            \State $\textit{start} \gets \textit{projection[0]}$ \Comment{Start of new region on projection axis}
        \EndFor
    \EndFunction
\end{algorithmic}
\end{algorithm}

\subsection{Energy}
The Otii power monitor returns the total energy absorbed during the execution of the experiment. Given that the duration of an experiment is not fixed, as we will see in Section~\ref{sec:results}, we normalized the measured energy over the experiment execution time. The result is the average power consumed by the device in the experiment.

\subsection{Pareto Efficiency}
\begin{figure}[!t]
    \centering
    \includegraphics[width=\columnwidth]{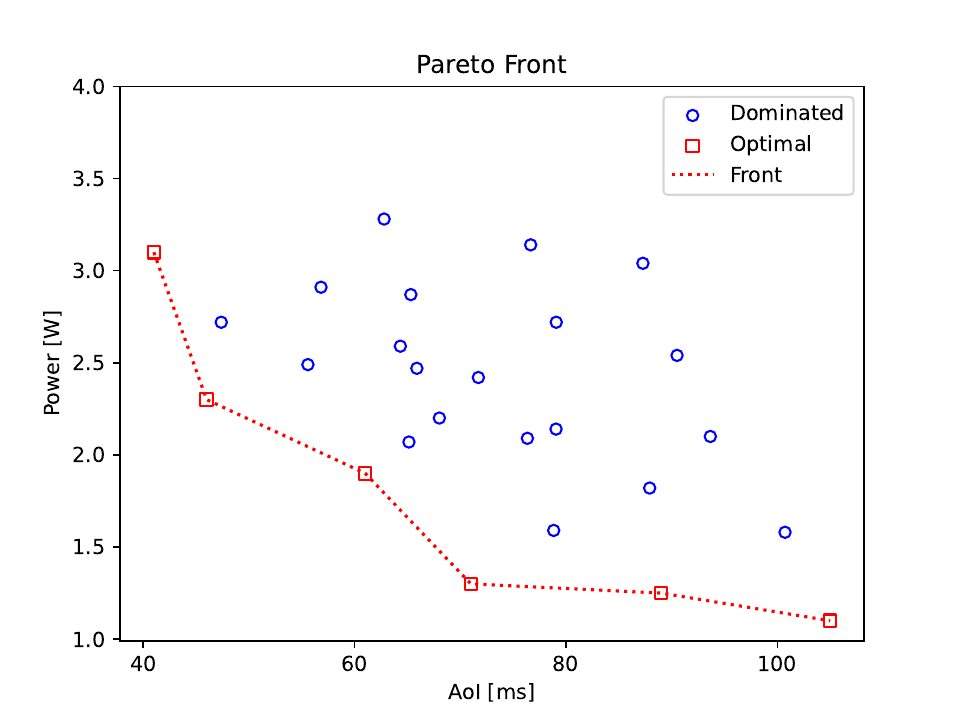}
    \caption{Example of Pareto Front.}
    \label{fig:pareto-example}
\end{figure}

The two metrics, \acrshort{aoi} and energy, are considered together for the energy-constrained nature of \acrshort{iot} devices. This can introduce additional complexity in the analysis: the objective is to optimize both, but there is a trade-off to be resolved. 
The best method for presenting the results is to use a tool that shows all possible optimal solutions and leaves it up to the reader to determine which of those presented is the most suitable configuration for a given system. The tool that perfectly matches the above requirements is the Pareto front, built up by a set of Pareto efficiencies. Pareto efficiency is a concept introduced by the Italian economist and engineer Vilfredo Pareto that states that, given a set of resources an allocation is efficient if it is not possible to improve the condition of one individual without worsening those of another, i.e. Pareto improvements are not available. This concept can also be applied in multi-objective optimization when there is no feasible solution that minimizes all objective functions simultaneously~\cite{Ngatchou2005:pareto}. Let us consider a set of points in the plane whose coordinates represent \acrshort{aoi} and energy, respectively, each associated with a particular configuration. The Pareto efficient solutions are those for which one metric cannot be improved without worsening the other. The sub-set of Pareto efficient solutions represents the Pareto front on which a designer can choose the solution to adopt, considering possible constraints of the system in terms of \acrshort{aoi} and energy. The choice can be considered as an a-posteriori assignment of weights to the metrics, based on the knowledge of the considered system. The points having both coordinates worse than other points are referred to as Pareto-dominated and the associated configuration doesn't provide an optimal choice for any system, thus they can be discarded. An example of the Pareto front is reported in Figure~\ref{fig:pareto-example}.

\section{Experimental Results}
\label{sec:results}

\begin{figure*}[!t]
\centering
\subfloat[Real rate vs. nominal rate.]{\includegraphics[width=0.32\textwidth]{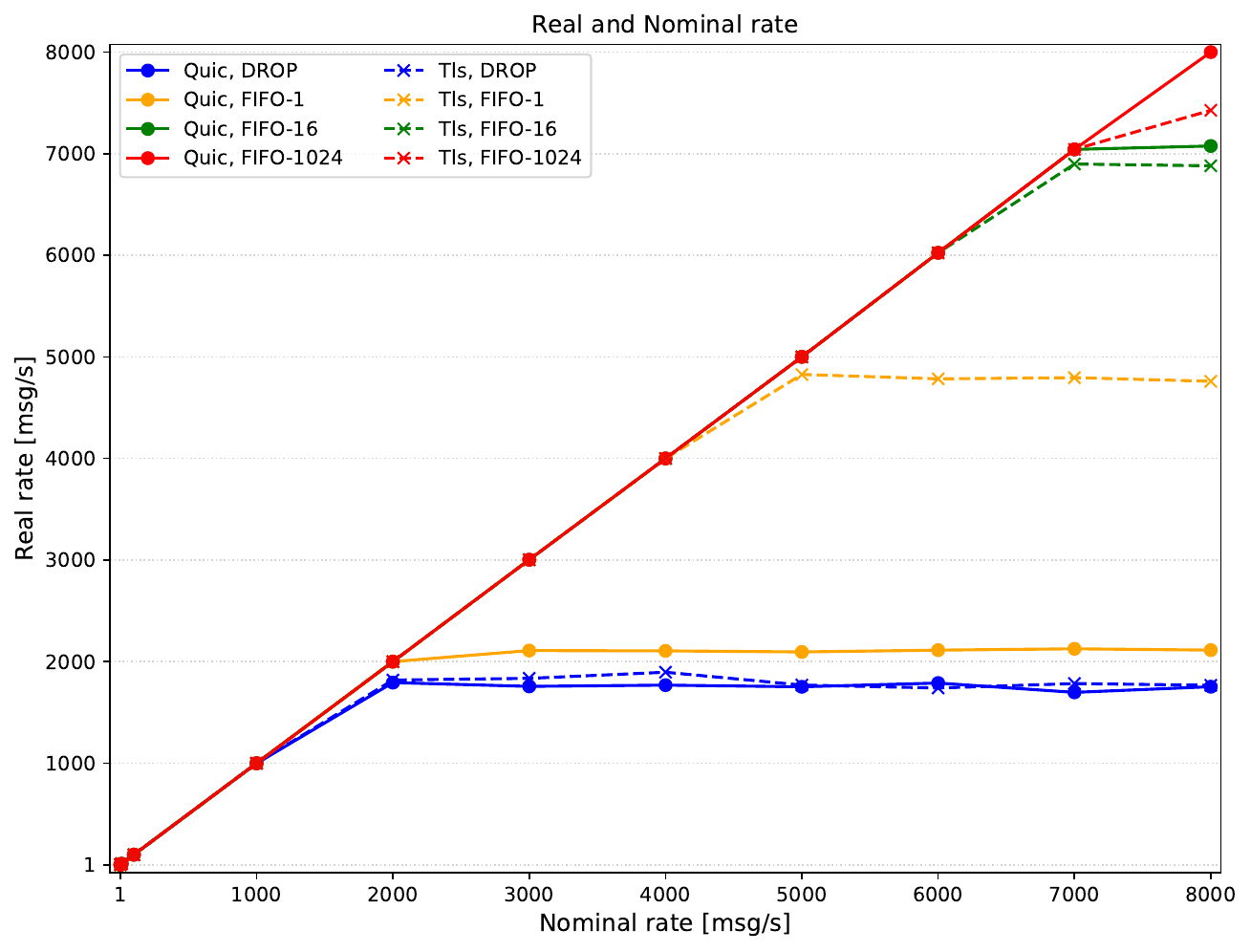} \label{fig:real-nominal-rate}}%
\subfloat[Median \acrshort{aoi} over rate]{\includegraphics[width=0.32\textwidth]{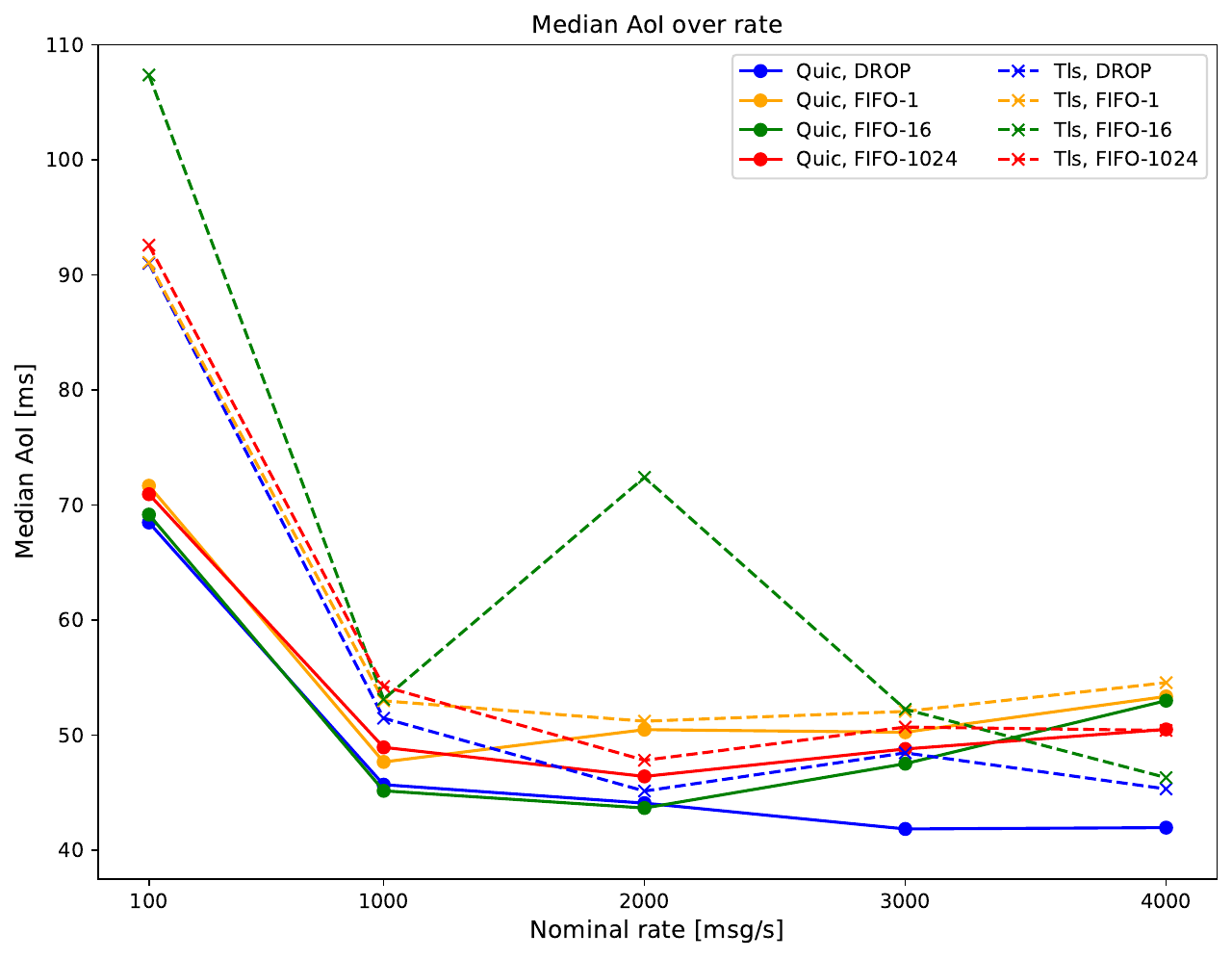} \label{fig:median-rate-medium}}%
\subfloat[Average power over rate]{\includegraphics[width=0.32\textwidth]{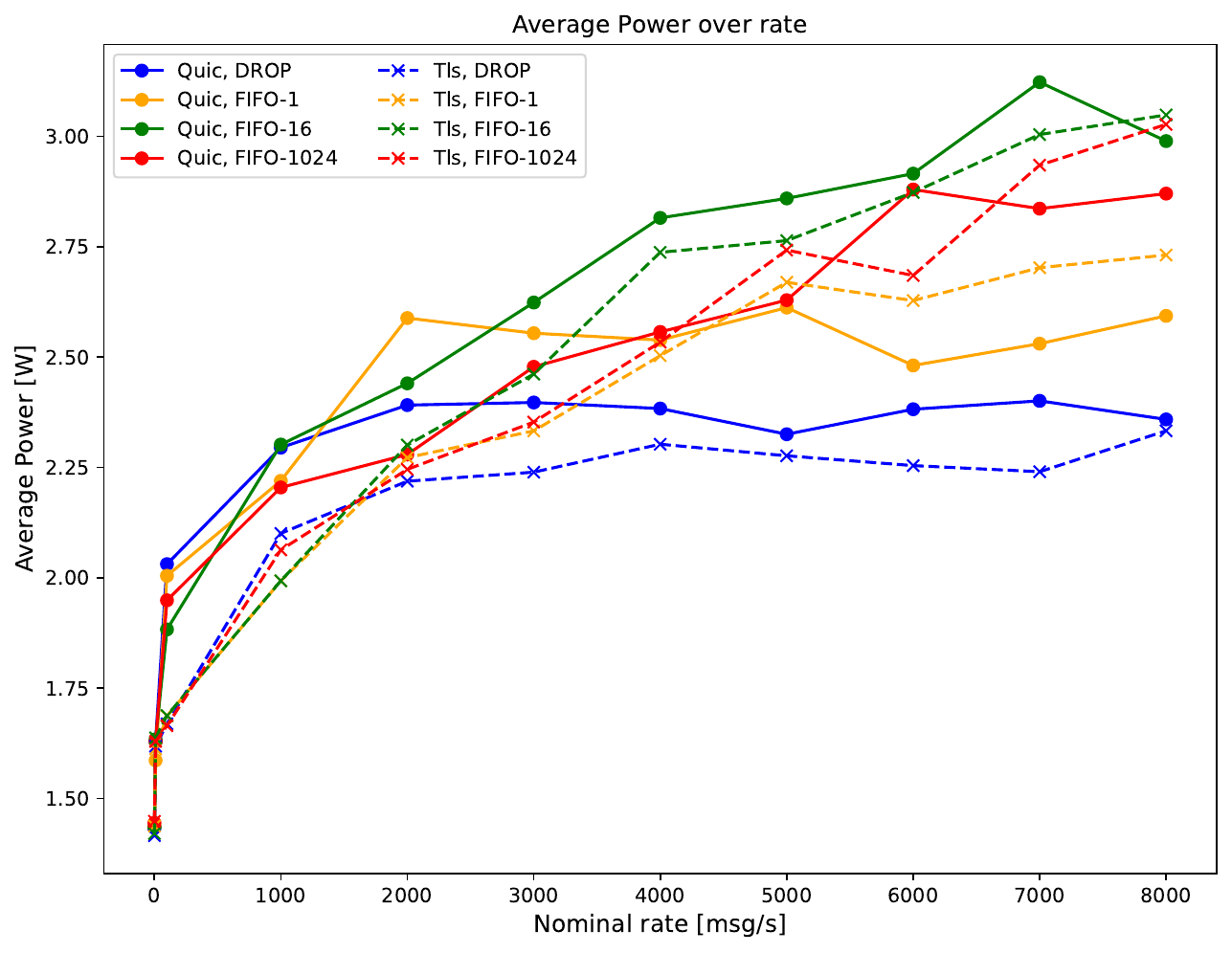} \label{fig:power-over-nominal-rate}}\\
\subfloat[Pareto front for FIFO configurations]{\includegraphics[width=.33\textwidth]{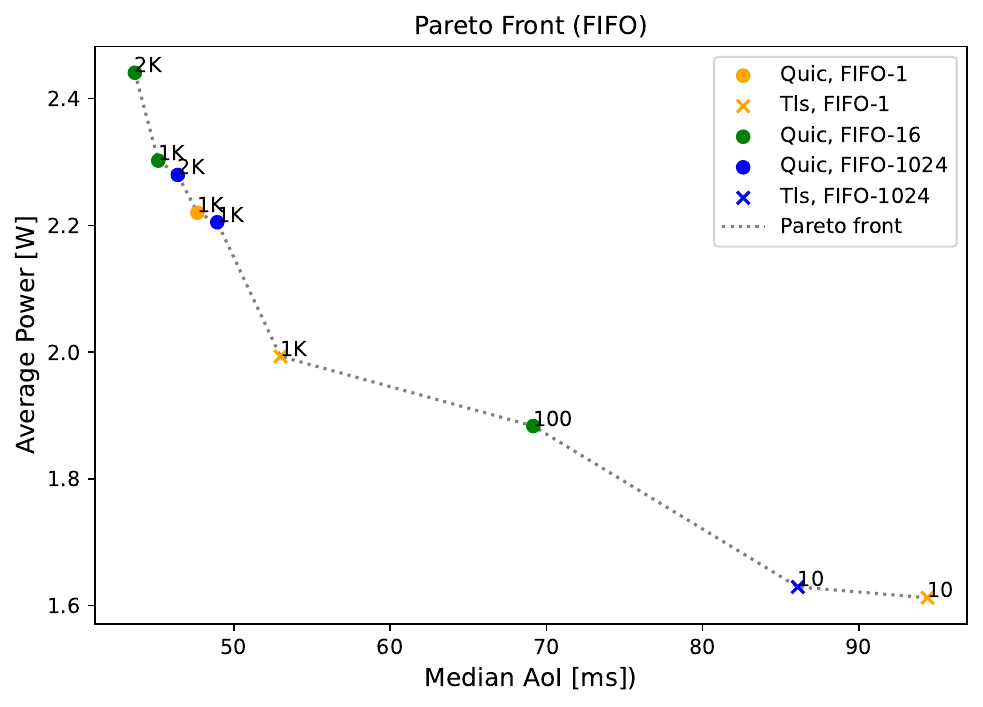} \label{fig:pareto-front-fifo-zoom}}%
\subfloat[Pareto front for DROP configurations]{\includegraphics[width=.33\textwidth]{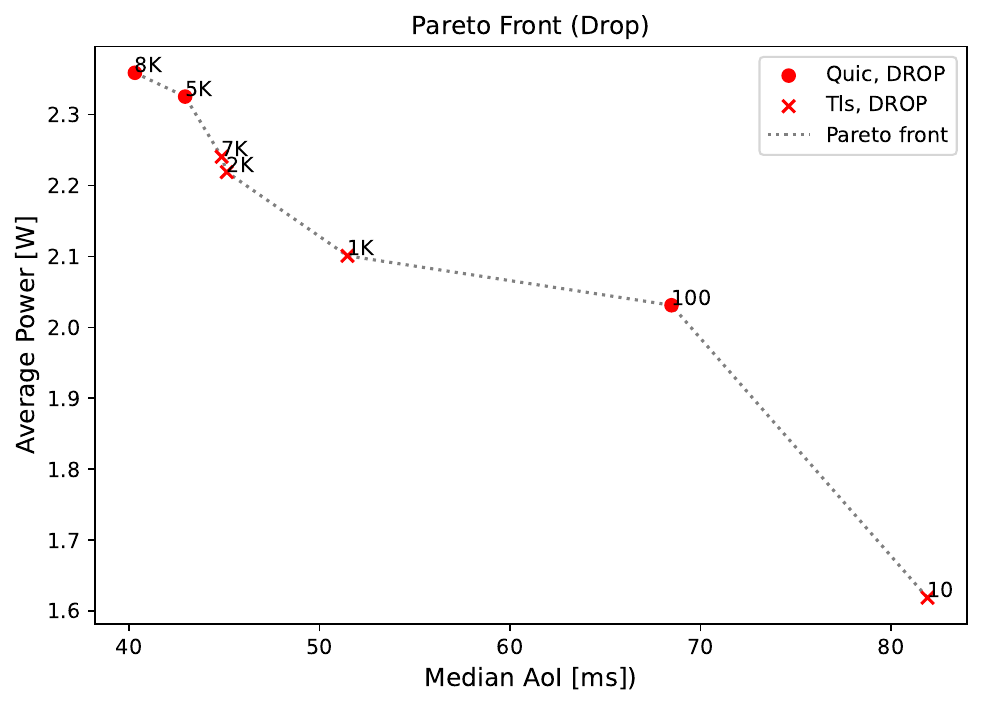} \label{fig:pareto-front-drop-zoom}}%
\caption{Generation rate results.}
\label{fig:rate}
\end{figure*}

Our experiments aim to evaluate the combination of the two metrics of interest when varying different operation parameters. Specifically, we evaluated the impact on energy and \acrshort{aoi} when varying: (i) the publisher generation rate, (ii) the additional delay between the publisher and the broker, i.e. the \acrshort{rpi} and the server, (iii) the payload size of the publisher messages, (iv) the number of cores used on the \acrshort{rpi}, (v) the transport protocol. The number of generated messages in one run is equal to the nominal generation rate multiplied by 60, which corresponds to a duration of 60 seconds. However, as we will highlight in the next sections, the real generation rate can be lower than the nominal rate. In these cases, the duration of an experiment will be longer. For each configuration, we executed one run, except for the last one, where we executed 30 runs. Besides the metrics used to compute energy and \acrshort{aoi}, the \acrshort{rpi} collects also context metrics useful to explain the obtained results, such as the cellular network quality and the clock synchronization precision.

\subsection{Impact of the Generation Rate}
For these experiments, we set the message payload to 128~B, no additional delay on the cellular link, and a single core on the \acrshort{rpi}. We varied the nominal generation rate on the publisher from 1 msg/s to 8000 msg/s, with the following steps: 1, 10, 100, 1000, 2000, ..., 8000. We show results for both the transport protocols (\acrshort{quic} and \acrshort{tls}), and all the buffering strategies. Figure~\ref{fig:real-nominal-rate} shows the relationship between the real generation rate and the nominal generation rate, for both transport protocols and the different buffering strategies. The figure shows that the buffer size poses an upper limit on the real generation rate. The bigger the buffer, the higher the real rate that can be obtained. For FIFO policies, this happens because, once the buffer is full, the publisher blocks. This automatically reduces the real publishing rate so that it is compatible with the maximum message throughput that can be sent out via the network. The presence of a larger buffer allows the publisher to cope with the transient reduction of throughput due to changing network conditions and, more importantly, it allows the aggregation of multiple messages within a single transport-layer packet. For the DROP buffering strategy, the real rate corresponds to the number of messages that are transmitted by the \acrshort{rpi}. It has to be noted that for the FIFO 1 policy, the rate obtained by \acrshort{tls} is much higher than the one obtained by \acrshort{quic}. This happens because in its base configuration, \acrshort{tcp} with \acrshort{tls} uses the Nagle algorithm, which performs a much more aggressive aggregation than \acrshort{quic}.

For each rate and each protocol/transport configuration, we computed median \acrshort{aoi} and average power. The median \acrshort{aoi} starts from high values when the rate is low, and then decreases to around the nominal delay value when the rate is 2000 msg/s. Then, it starts increasing again for all FIFO configurations. For the DROP configurations, instead, it remains at the minimum value. This is expected, as the property of the DROP policy is to deliver always the most ``fresh'' message. Figure~\ref{fig:median-rate-medium}, shows this trend for rates from 100 to 4000. It must be noted that the \acrshort{quic} configurations obtain always lower \acrshort{aoi} values than their corresponding \acrshort{tls} ones. Figure~\ref{fig:power-over-nominal-rate} shows the power consumption over the nominal rate. After this initial sudden increase, the power continues rising with the rate but with a slower slope. This trend is valid for all configurations up to the rate of 2000 msg/s. From such a rate onward the DROP policy stops increasing and the same applies for FIFO 1, but only in the \acrshort{quic} configuration. The FIFO-1 \acrshort{tls} encounters an upper bound around the rate of 5000 msg/s. Other configurations, even if subjected to some fluctuations, present a quite evident increasing trend up to the maximum rate. This suggests that the power consumption is dominated by the transmission rate. Once the maximum rate for a configuration is reached, further increases in the nominal rate won't produce any significant variation in the average power. It must be noted that below 2000 msg/s \acrshort{quic} always shows a higher power consumption. Over that threshold, things get more confused, even if generally \acrshort{quic} consumes more than its corresponding \acrshort{tls} configuration.

Figures~\ref{fig:pareto-front-fifo-zoom} and~\ref{fig:pareto-front-drop-zoom} show the Pareto fronts for the FIFO policies and the DROP policies, respectively. The results for the rate of 1 msg/s are omitted for the sake of clarity, as they obtain a median \acrshort{aoi} of approximately 500 ms, even if the power consumption is extremely low. For the FIFO configurations, the rates that appear in the front do not exceed 2000 msg/s. As previously observed, excessively high transmission rates do not guarantee advantages for either power consumption or \acrshort{aoi}. Besides that, in general, the lower the rate the lower the consumption and the higher the \acrshort{aoi}, and vice-versa, and the \acrshort{quic} configurations obtain the lower \acrshort{aoi}. It is worth noticing that the high density of points in the left extreme also shows that, once approaching the lower \acrshort{aoi} bound, any further improvement requires a significant increment in the power consumption. In the DROP configurations, we can notice that also rates higher than 2000 msg/s are present in the front. This happens because higher rates, even if higher than the nominal one, produce ``fresher'' information, thus helping keep the \acrshort{aoi} low, at the cost of possible information loss.

\begin{figure}[!t]
    \centering
    \includegraphics[width=\columnwidth]{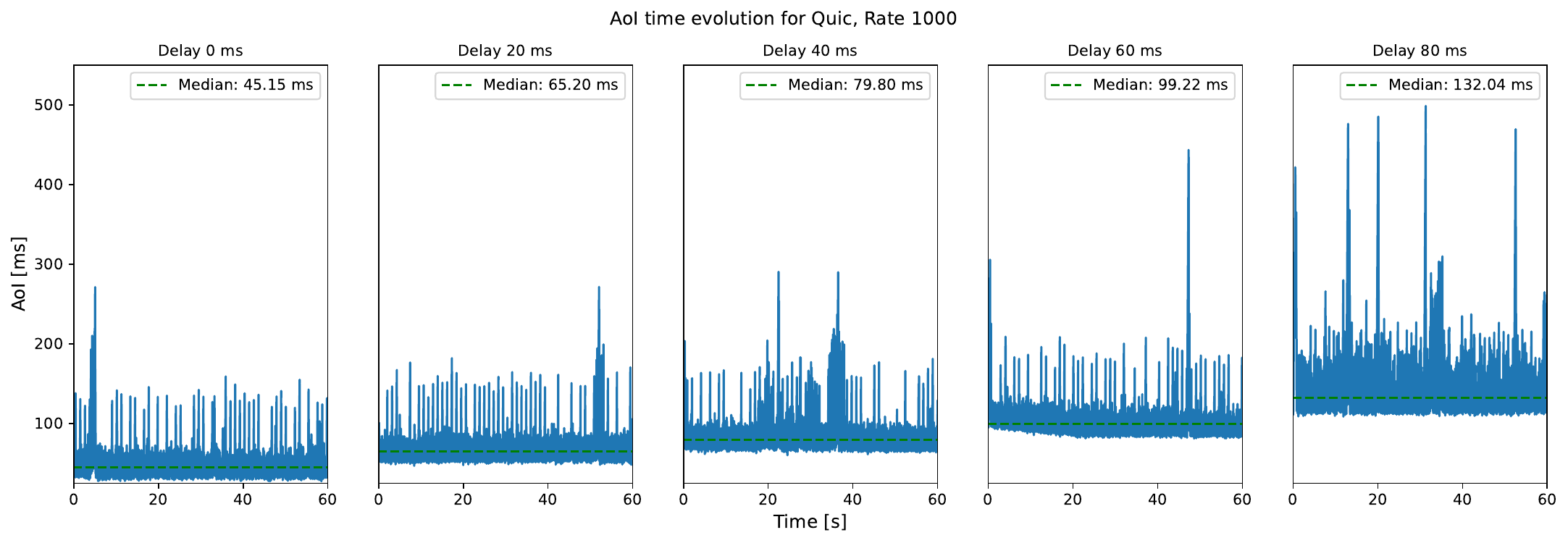}
    \caption{\acrshort{aoi} values for different additional delays.}
    \label{fig:aoi-time-evolution-delay-comparison}
\end{figure}

\begin{figure*}[!t]
\centering
\subfloat[Maximum rate for different delays.]{\includegraphics[width=0.32\textwidth]{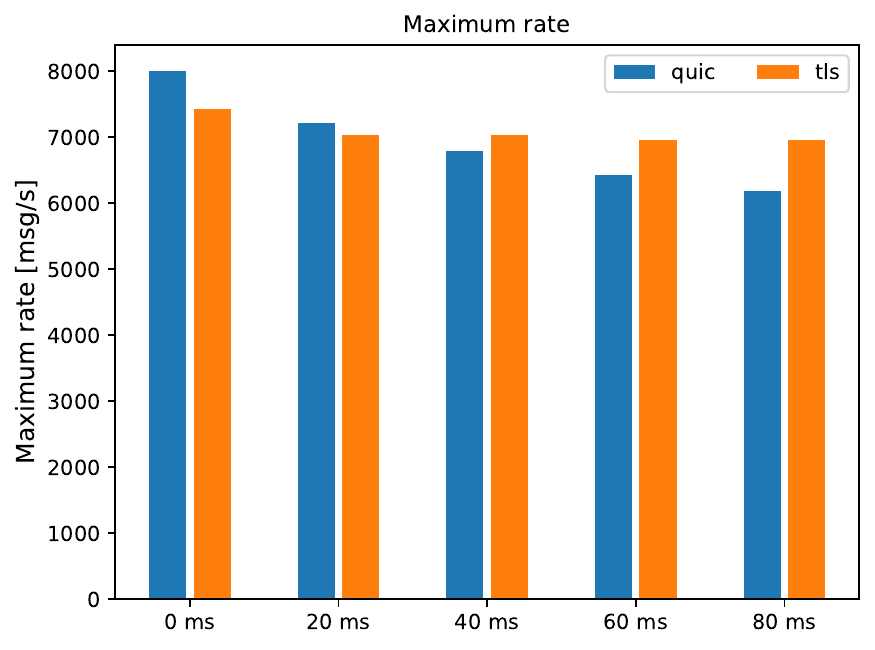} \label{fig:maximum-rate-delay}}%
\subfloat[Pareto fronts for FIFO delayed configurations]{\includegraphics[width=0.32\textwidth]{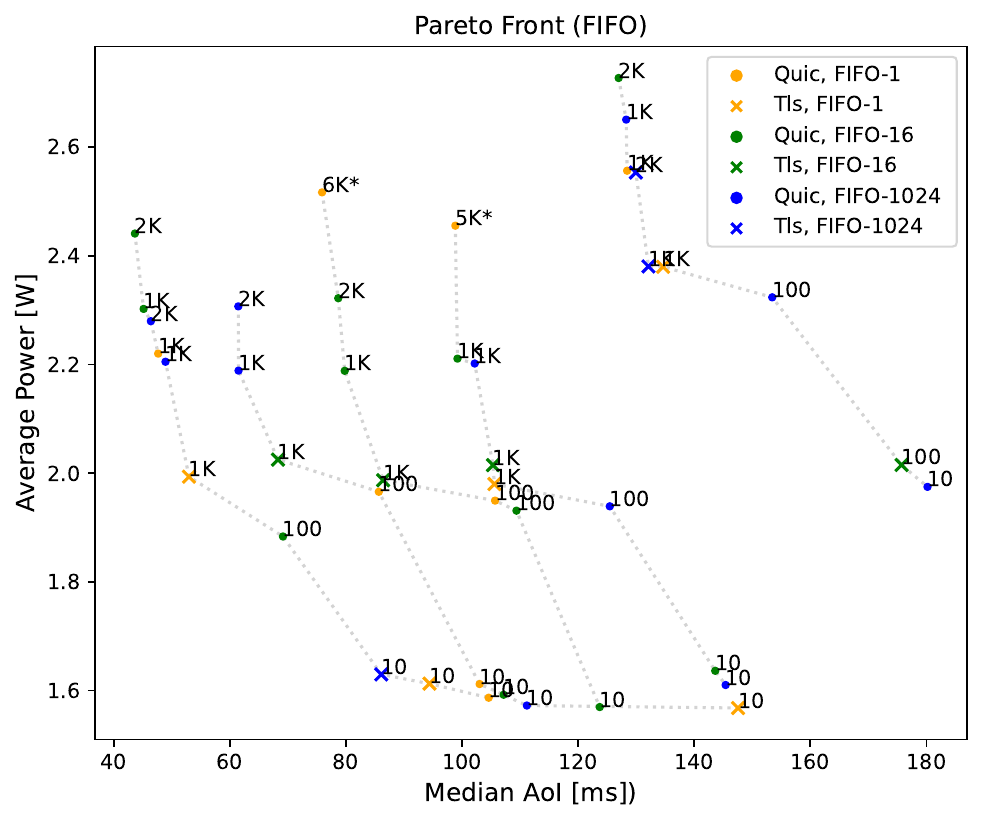} \label{fig:pareto_front_delay_fifo}}%
\subfloat[Pareto fronts for DROP delayed configurations]{\includegraphics[width=0.32\textwidth]{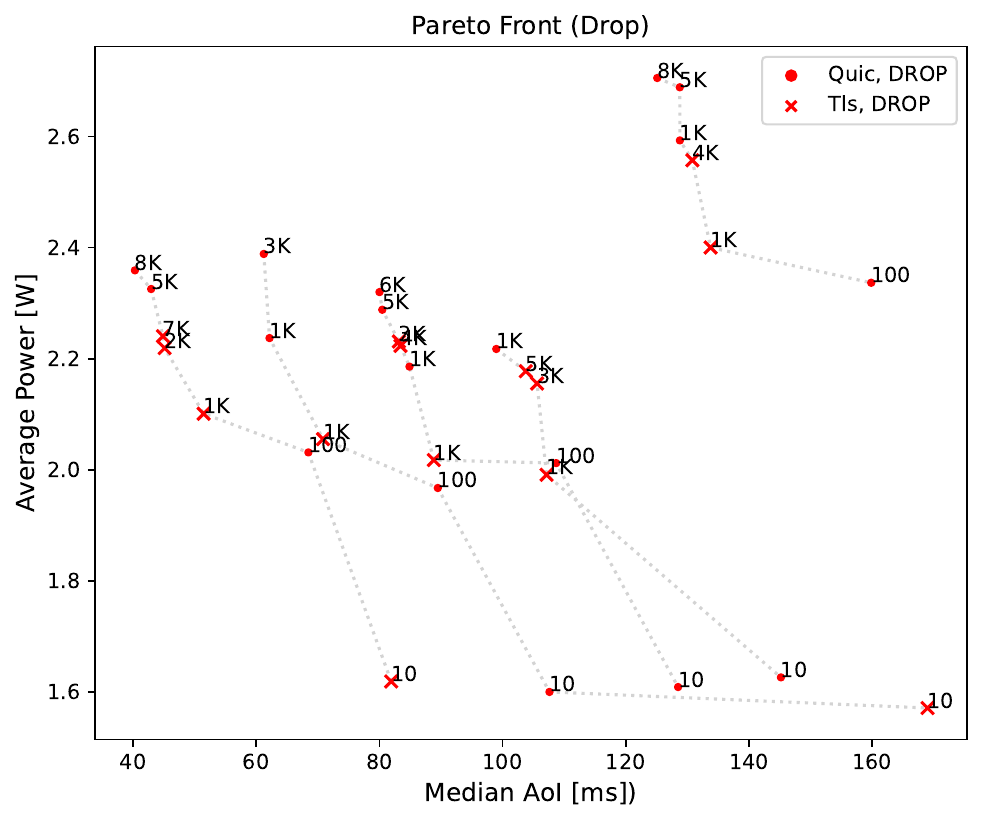} \label{fig:pareto_front_delay_drop}}
\caption{Additional delay results.}
\label{fig:delay}
\end{figure*}

\subsection{Impact of Additional Delay}
For these experiments, we artificially applied additional delay on the cellular link in both directions (uplink and downlink). Specifically, the delay was added on the server machine via the Linux \acrfull{tc} command. We performed runs with additional delays from 20 to 80 ms, at steps of 20 ms. For what concerns \acrshort{aoi}, as could be expected, an upward shift can be appreciated in Figure~\ref{fig:aoi-time-evolution-delay-comparison}, which shows the raw results for the QUIC, FIFO 16 configuration with a rate of 1000 msg/s. The shift is less evident for lower rates, as the \acrshort{aoi} values are already high. The maximum achievable rate is affected too. In general, a long queue allows higher rates, as already observed. However, for longer queues (namely FIFO 16 and FIFO 1024), by increasing the delay there is an overall reduction of the maximum rate, mainly for QUIC. The phenomenon can be visualized in Figure~\ref{fig:maximum-rate-delay}, where, starting from a 40 ms additional delay, the rate achieved by \acrshort{quic} is progressively lower than the one obtained by \acrshort{tls}. The causes for this behavior are beyond the scope of this work, however, they could lie at the intersection of the following aspects:
\begin{itemize}
    \item \textbf{Congestion control}. The congestion control algorithm is the same, Cubic, for both transport protocols. However, the operational parameters may differ, resulting in different performance when varying the network delay.
    \item \textbf{Flow control}. The advanced flow control mechanism of \acrshort{quic} based on multiple streams is not exploited by the implemented tool, so in this configuration also \acrshort{quic} may suffer from the head-of-line blocking problem. 
    \item \textbf{Aggregation}. The aggregation at the transport layer for \acrshort{tls}/\acrshort{tcp} is based on the Nagle algorithm, that, for high transmission rates, allows for sending only fully sized segments, independently from the network delay, while \acrshort{quic} adopts a blander aggregation mechanism.
    \item \textbf{Kernel optimization}. The Linux kernel on the \acrshort{rpi} doesn't offer optimization for the \acrshort{udp} protocol, such as \acrfull{gso}, potentially reducing the performance of \acrshort{quic}.
\end{itemize}
We conclude the analysis of the impact of the additional delay by showing the Pareto fronts for FIFO and DROP buffering strategies (Figures~\ref{fig:pareto_front_delay_fifo} and~\ref{fig:pareto_front_delay_drop}, respectively). In the figures, for each additional delay (including no additional delay) a separate front is depicted. The leftmost front is the one with the minimum delay and the rightmost front is the one with the highest delay. For the FIFO configurations, the considerations for the case without additional delay apply also to the other cases, i.e. lower rates guarantee a low power consumption, at the cost of higher \acrshort{aoi}, and vice-versa. In addition, \acrshort{quic} configurations obtain lower \acrshort{aoi}, while \acrshort{tls} ones obtain lower energy consumption, as already observed in the case with no additional delay. Rates higher than 2000 msg/s do not guarantee any benefits except in two cases. In both cases, these results are obtained with a FIFO 1 queue and the \acrshort{quic} protocol. Considering that in the FIFO configurations, the generation rate adapts to the transmission rate, the result is to be considered equal to one obtained by a lower rate. The improvement in terms of \acrshort{aoi} with respect to the second highest point in the front is negligible, at the unjustifiable cost of higher energy consumption. The same considerations found for the case with no additional delay apply also to the DROP configurations. In particular, these configurations benefit also from higher rates, as already pointed out.

\begin{table}
    \centering
    \caption{Median \acrshort{aoi} varying the payload size for \acrshort{tls}, FIFO-16, rate 100 configuration}
    \label{tab:aoi-payload-tls-q16}
	\begin{tabular}{lr}
		\toprule\toprule
		\multicolumn{1}{c}{\textbf{Payload size}} & \multicolumn{1}{c}{\textbf{Median \acrshort{aoi}}} \\
		\midrule
		128B & 107.38 \\
		\midrule[0pt]
        256B & 85.04  \\
		\midrule[0pt]
		512B & 65.52  \\
        \midrule[0pt]
		1KB & 60.24  \\
        \midrule[0pt]
		2KB & 55.46  \\
		\bottomrule
	\end{tabular}
\end{table}

\begin{figure*}[!t]
\centering
\subfloat[FIFO.]{\includegraphics[width=0.48\textwidth]{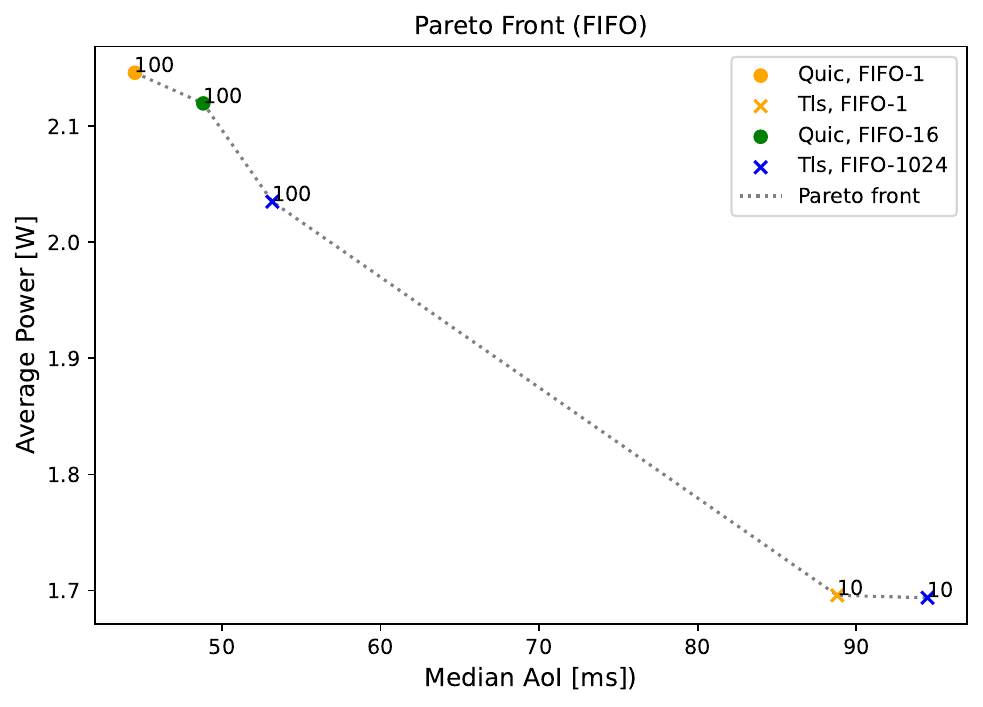} \label{fig:pareto-front-2kb}}%
\subfloat[DROP.]{\includegraphics[width=0.48\textwidth]{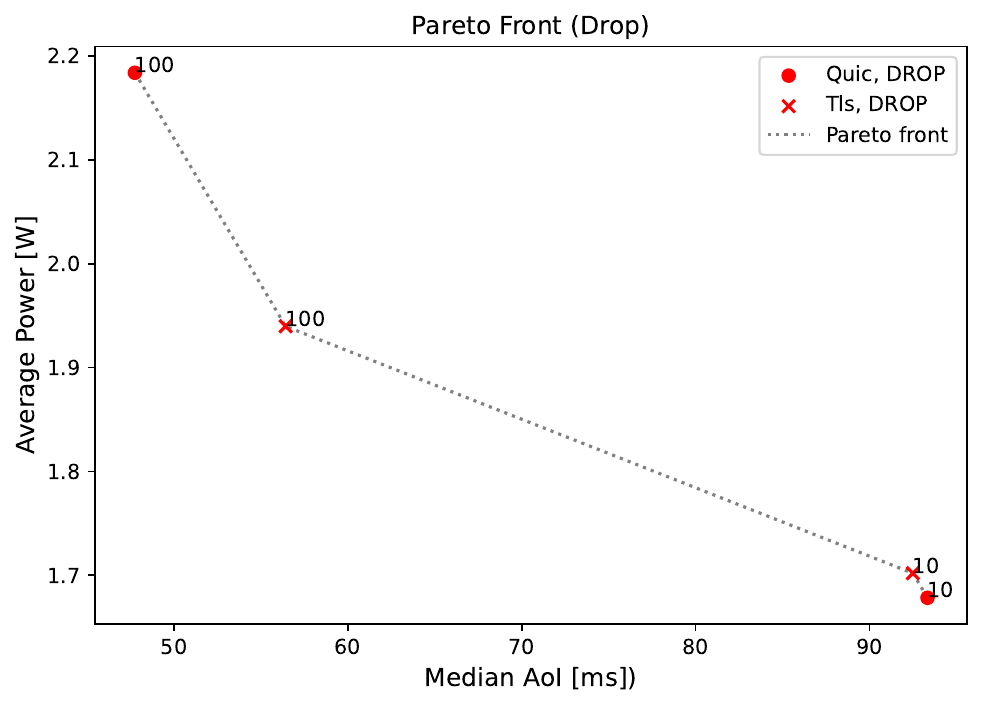} \label{fig:pareto-front-2kb-drop}}%
\caption{Pareto fronts for 2048~B payload configurations.}
\label{fig:pareto-payload}
\end{figure*}

\begin{figure*}[!t]
\centering
\subfloat[FIFO 1024.]{\includegraphics[width=0.48\textwidth]{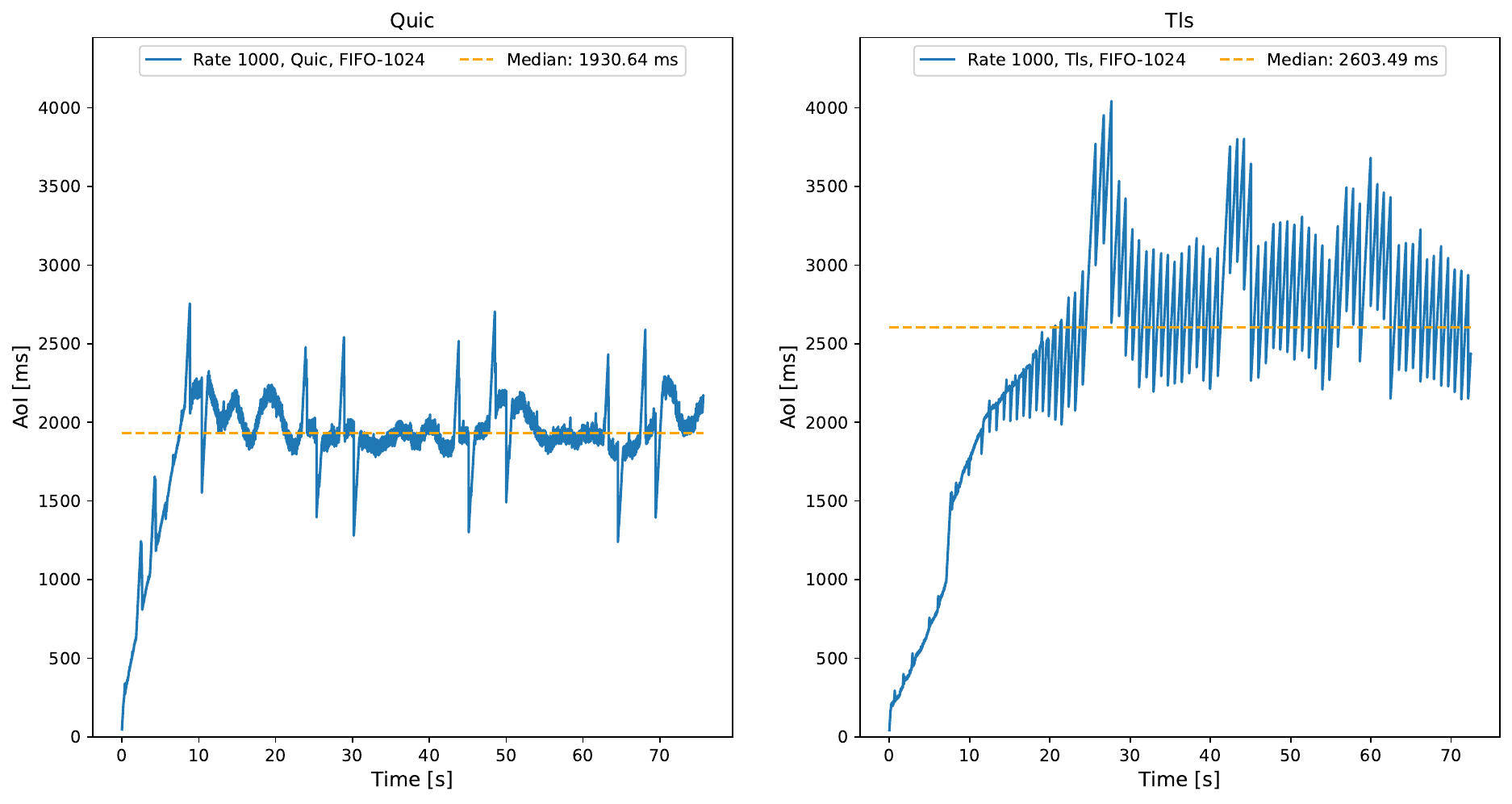} \label{fig:aoi_time_evolution_2KB_Q1024_R1000}}%
\subfloat[DROP.]{\includegraphics[width=0.48\textwidth]{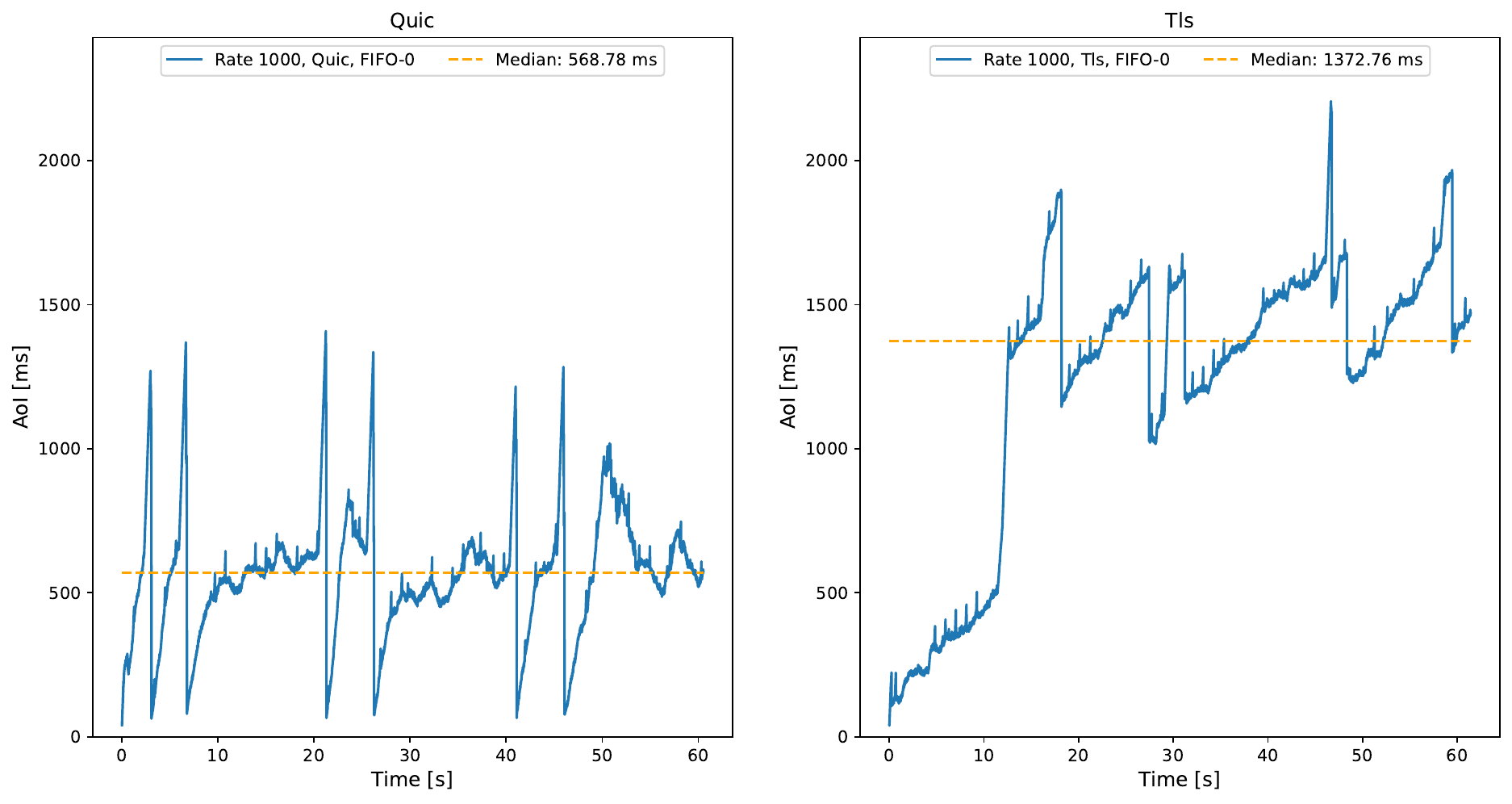} \label{fig:aoi_time_evolution_2KB_Q0_R1000}}%
\caption{Time evolution for 2048~B payload, rate 1000 msg/s.}
\label{fig:delay-time-evo}
\end{figure*}

\subsection{Impact of Different Payload Sizes}
In previous experiments, we considered a fixed payload size of 128~B. However, there could be scenarios characterized by advanced sensors producing more complex, thus bigger data structures. In addition, multiple sensors could be attached to the same device which would in turn produce bigger packets containing more than one message. In these experiments, we evaluate the impact of greater payload sizes on energy consumption and \acrshort{aoi}. We consider the following payload sizes: 128~B, 256~B, 512~B, 1024~B, 2048~B. The maximum payload size of 2048~B allows us to consider application scenarios where a large amount of data has to be fitted in a single message, as well as to account for possible effects on the \acrshort{aoi} of messages that do not fit in a single segment at the transport layer. For these experiments, we considered only rates up to 1000 msg/s as (i) the best results for \acrshort{aoi} in the previous experiments were obtained with rates around 1000-2000 msg/s, (ii) the combination of large payload size and high transmission rate would produce a very high throughput witch could not be sustainable by the \acrshort{rpi} and \acrshort{iot} devices in general. The analysis of the Pareto fronts highlights a slightly higher power consumption for the configurations with higher payload sizes, while the \acrshort{aoi} decreases as the payload increases, for the rate of 100 msg/s and in some configurations even 1000 msg/s. Table~\ref{tab:aoi-payload-tls-q16} shows this behavior for the \acrshort{tls}, FIFO 16, rate 100 msg/s configuration. Particularly interesting are the Pareto fronts for the payload size of 2048~B (Figure~\ref{fig:pareto-payload}), which show the configurations with a rate of 100 mgs/s as the best ones for \acrshort{aoi}. In fact, at higher rates, the \acrshort{aoi} values are extremely high. This happens because the system is not able to handle the required throughput, and this does not depend on the queue size, as testified by Figure~\ref{fig:delay-time-evo}, which shows the anomalous behavior for both the FIFO 1024 and the DROP queues. Therefore, the cause seems to be due to oversized buffering that takes place somewhere between the transmitter and the receiver, either in the \acrshort{rpi} (on-device bufferbloat~\cite{bufferbloat}) or in the cellular network infrastructure. It has to be noticed a difference between the \acrshort{aoi} obtained with \acrshort{quic} and \acrshort{tls}, with \acrshort{quic} that always obtains a better performance. For what concerns power consumption, the least power-demanding transmission protocol is \acrshort{tls}, however, when the payload size increases the difference between the protocols tends to shrink, especially for the FIFO configurations.

\begin{figure*}[!t]
\centering
\subfloat[\acrshort{aoi} comparison enabling second core for \acrshort{quic}, FIFO-1.]{\includegraphics[width=0.35\textwidth]{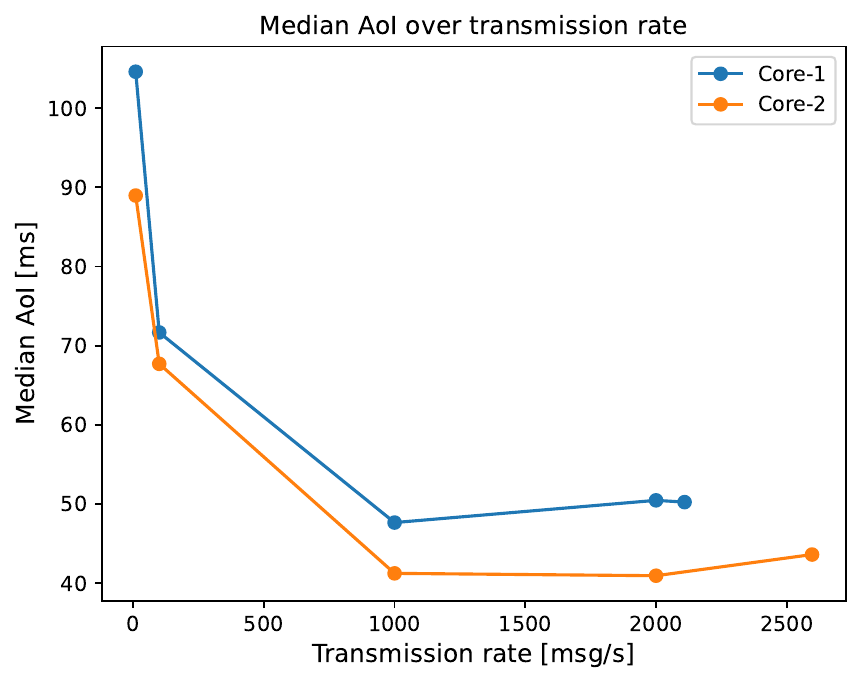} \label{fig:aoi_cores_comparison}}%
\subfloat[Energy comparison enabling second core for \acrshort{quic}, FIFO-1 and DROP.]{\includegraphics[width=0.64\textwidth]{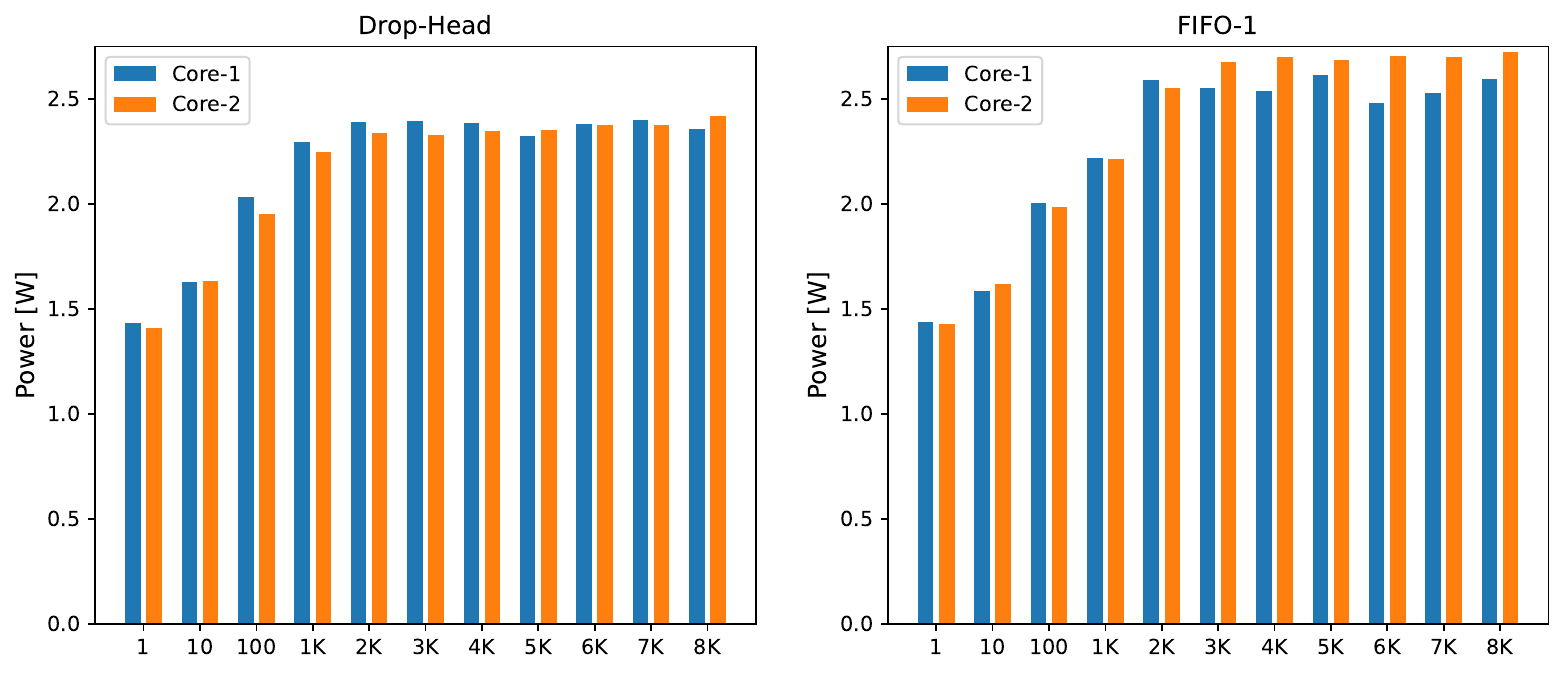} \label{fig:energy_cores_comparison}}%
\caption{Results for multiple cores configurations.}
\label{fig:cores}
\end{figure*}

\subsection{Impact of Multiple Cores}
In this experiment, we evaluate the impact on power consumption and \acrshort{aoi} of enabling an additional core on the \acrshort{rpi}. The software we developed is single-threaded, and concurrency is implemented via asynchronous mechanisms, that exploit the idle times dictated by I/O operations to carry out other tasks. However, in parallel with our software, multiple kernel threads are executed, responsible for handling network operations and other system tasks. Thus, enabling more than one core could benefit the overall system's performance.

\begin{figure*}[!t]
\centering
\subfloat[\acrshort{aoi} over transmission rate.]{\includegraphics[width=0.48\textwidth]{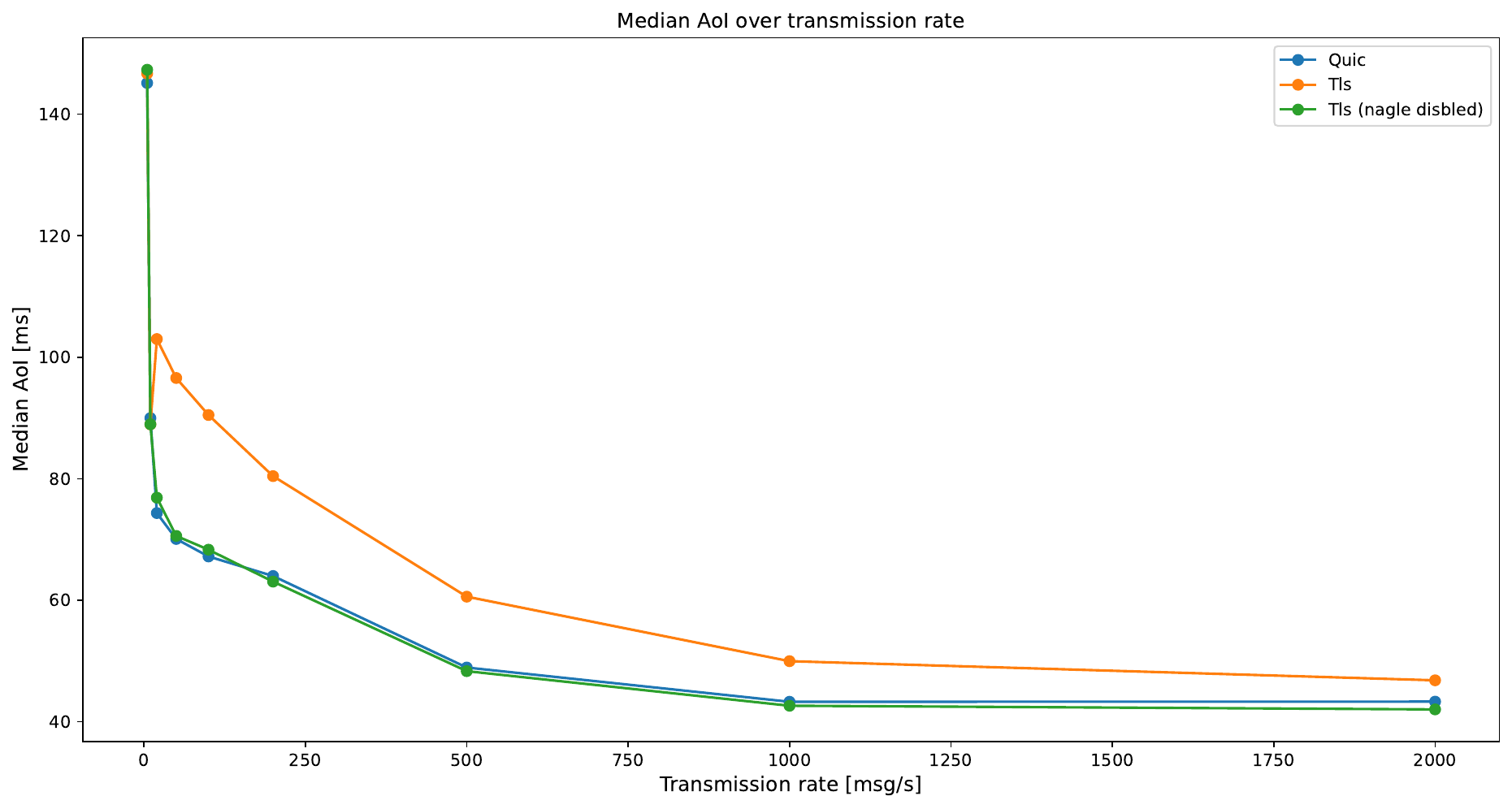} \label{fig:aoi-fifo-repetition}}%
\subfloat[Power over transmission rate.]{\includegraphics[width=0.48\textwidth]{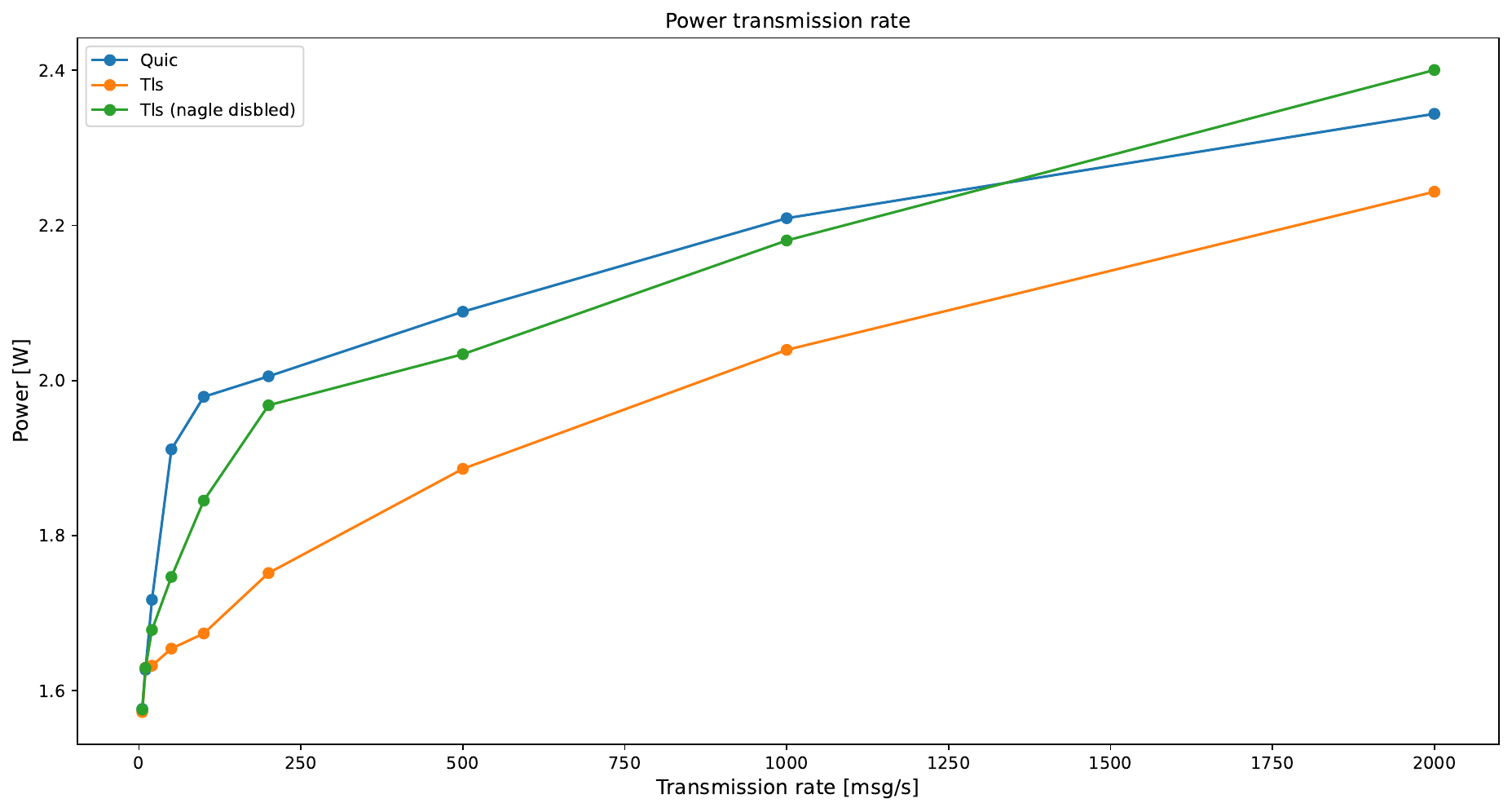} \label{fig:power-fifo-repetition}}%
\caption{Impact of the transport protocol for FIFO 16 configurations.}
\label{fig:transport-protocols}
\end{figure*}

Enabling the second core allowed the system to reach higher rates, and, more importantly, a lower \acrshort{aoi}. This is particularly evident for \acrshort{quic} in the FIFO 1 configurations, as can be observed in Figure~\ref{fig:aoi_cores_comparison}. This effect is not found in the DROP configurations, as the possibility to overwrite the buffer content allows always to deliver the ``freshest'' information. The positive impact of the second core has a drawback, paid in terms of additional energy required to power the hardware component, but not in all configurations, as shown in Figure~\ref{fig:energy_cores_comparison}. For DROP configurations, it is not possible to spot a clear difference in power consumption. For FIFO configurations instead, the impact on the power consumption is significant as the rate gets higher. This is because the second core allows higher transmission rates, which, in turn, can increase the power consumption of the network card.

\subsection{Impact of the Transport Protocol}
We finally executed a set of experiments aimed at quantifying the impact of the transport protocol on power consumption and \acrshort{aoi}. We considered three transport protocols: \acrshort{quic}, \acrshort{tls}, and \acrshort{tls} with Nagle's algorithm disabled. We added the latter because, as aforementioned, the aggregation plays a crucial role in the maximum achievable rate and the \acrshort{aoi}, often resulting in a discriminant factor between \acrshort{quic} and \acrshort{tls}. We considered just two buffering policies: FIFO 16 and DROP. We used just a single core and a fixed payload size of 128~B. We limited the maximum generation rate to 2000 msg/s but we explored more values in the range of generation rates, in particular: 5, 10, 20, 50, 100, 200, 500, 1000, 2000. No additional delays were considered. Finally, we kept the duration of a single experiment to 60 s, but we repeated each run 30 times. This is to ensure statistical robustness. We aggregated the results (power consumption and \acrshort{aoi}) of each repetition of the same setup by taking the median value. The results for the two buffering strategies are extremely similar, thus, to avoid confusion, we show just the ones for FIFO 16.

Figure~\ref{fig:aoi-fifo-repetition} shows the \acrshort{aoi} obtained when varying the transmission rate. The better performance in terms of \acrshort{aoi} of \acrshort{quic} with respect to standard \acrshort{tls} is confirmed. However, the most interesting result is that, by disabling Nagle's algorithm, \acrshort{tls} yields the same if not slightly better results as \acrshort{quic}. This result shows that the main difference in terms of \acrshort{aoi} between the transport protocols is given by the aggregation mechanism at the transport layer, much more aggressive when Nagle's algorithm is enabled. Standard \acrshort{tls} shows a peculiar trend for rates higher than 10 msg/s. \acrshort{aoi} suddenly and unexpectedly increases for a 20 msg/s rate, to start decreasing again afterward, but assessing of significantly higher values than the other two protocols. This is due to Nagle's algorithm, which is triggered by acknowledgments: when the transmission rate produces messages with a period higher than the network RTT (which we recall is $\sim$80 ms between the publisher and the broker), Nagle's algorithm delays transmission until acks are received or the segment gets full, and this negatively affects \acrshort{aoi} despite the increase of the transmission rate.

The \acrshort{aoi} improvement obtained by removing the aggregation has a consequence from the energy point of view, as shown in Figure~\ref{fig:power-fifo-repetition}. The \acrshort{tls} curve with Nagle's algorithm disabled is much closer to the \acrshort{quic} curve than the standard \acrshort{tls} curve. The \acrshort{quic} configuration seems to be the most energy-demanding, at least up to the highest considered rate, for which the curves swap. The lack of aggregation in the \acrshort{tls} with Nagle's algorithm disabled may cause the inversion. The \acrshort{quic} protocol is responsible for lighter aggregation than Nagle's algorithm, so the effects on the \acrshort{aoi} are limited, but when the transmission rate increases, having some messages shipped in the same transport layer packet allows for fewer transmissions and therefore less energy spent. A simple analysis of traces collected for a rate of 2000 msg/s shows a mild aggregation for \acrshort{quic}, around 2 messages per segment, while \acrshort{tls} without Nagle's algorithm doesn't provide any form of aggregation.

\begin{figure}[!t]
    \centering
    \includegraphics[width=0.85\columnwidth]{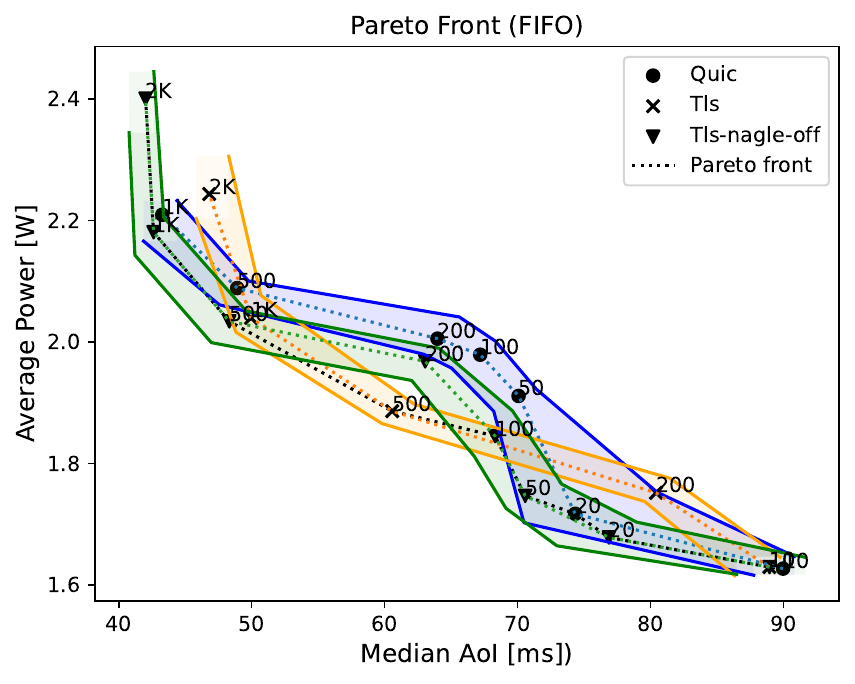}
    \caption{Pareto fronts for FIFO configurations.}
    \label{fig:pareto-fifo-repetitions}
\end{figure}

Figure~\ref{fig:pareto-fifo-repetitions} shows the Pareto fronts for the three considered protocols. Besides the median values, the $25^{th}$ and $75^{th}$ percentiles for both \acrshort{aoi} and power are also reported, which identify three bands around the Pareto front of each protocol. A larger band means a greater variability of data and thus less confidence in the single point obtained via aggregation. When the bands are far enough it is quite safe to state that one solution is better than the other, however, if the bands intersect each other, the two solutions are almost equivalent, thus the choice should be taken considering other aspects. The Pareto front identified by the combination of the three fronts is finally depicted as a black dotted line.

In the right part of the graph, the three transport protocols behave similarly, as highlighted by the three overlapped bands. Moving towards the left, \acrshort{tls} dominates the other protocols, leaving a space from one band to the others and having the Pareto front completely in its band. Keeping moving in the same direction the standard \acrshort{tls} is no longer able to guarantee the optimal results, so it is replaced by the other two protocols, for which there isn't a clear winner. It is worth noticing that \acrshort{quic} presents a larger variability, indicated by a larger band, especially for slower rates, i.e. on the right part of the graphs, while both versions of \acrshort{tls} generate a narrower band. Moreover, the Pareto fronts show that the aggregation of Nagle's algorithm can save energy, at the cost of a higher \acrshort{aoi}. The result is that for a middle region, where the \acrshort{aoi} is not minimized, the standard \acrshort{tls} seems to guarantee the best trade-off. When the requirements on \acrshort{aoi} are the most stringent, the aggregation limits the performance, thus disabling Nagle's algorithm or adopting \acrshort{quic} could help to meet the requirements. The simple fine-tuning of the \acrshort{tls} protocol, derived from the underlying \acrshort{tcp} option, can guarantee optimal protocol results even in those scenarios where \acrshort{quic} seemed to provide better results, so in many cases the greater availability of \acrshort{tls} may favor this protocol over the emerging \acrshort{quic}, for which the standardization process and availability are not yet comparable to the classic \acrshort{tls} over \acrshort{tcp}.
\section{Conclusion}
\label{sec:conclusion}

In this work, we conducted a combined study of \acrshort{aoi} and energy consumption in an \acrshort{iot} environment based on \acrshort{mqtt}, where a battery-operated device acts as a data publisher. Experimental results show that improving the AoI requires to increase the energy expenditure, as both of them depend somehow on the transmission rate. However, our contributions are not limited to this result. We considered a wide range of system operational parameters and showed their impact on both \acrshort{aoi} and energy. Our results provide system designers with guidelines and numbers useful for finding the desired trade-off between AoI and energy consumption, according to the system requirements.

Thanks to Pareto fronts, we highlighted all the optimal solutions, leaving to the final designer the possibility to choose which operational point better satisfies the specific system requirements. The results showed that a DROP buffering policy is better for the application queue management, but paying the cost of message loss which may not be admitted in some scenarios. Therefore, \acrshort{fifo} buffering strategies should be considered as well, especially when employing long queues, that allow for aggregation at the transport level resulting in a dramatic improvement in terms of reachable throughput. The aggregation at the transport layer was revealed to be a paramount factor to take into consideration. The aggressive aggregation enforced by Nagle's algorithm is the main factor causing different results for the two transport protocols, \acrshort{tls} and \acrshort{quic}. \acrshort{quic} able to guarantee a better \acrshort{aoi} than standard~\acrshort{tls} at the cost of higher energy consumption, at least for the rates employs in the analysis. This difference disappears when disabling Nagle's algorithm, making the two protocols comparable.

Overall, our study confirmed the possibility of adopting \acrshort{quic} as a reliable transport protocol, replacing \acrshort{tcp}. However, the immaturity of the new protocol may represent an obstacle to its adoption, favoring a more standardized protocol for which compatibility and support are much larger.

\section*{Acknowledgment}

\bibliographystyle{IEEEtran}
\bibliography{bibliography}

\end{document}